\documentclass[twocolumn,floatfix,amsmath,amssymb]{revtex4}
\usepackage{graphicx}
\usepackage{dcolumn}
\usepackage{color}

\newcommand{\be}{\begin{equation}}
\newcommand{\ee}{\end{equation}}

\def\ie{{\it i.e.}\ }

\newcommand{\bu}{\mathbf{u}} 
\newcommand{\bv}{\mathbf{v}}
\newcommand{\bw}{\mathbf{w}}

\begin{document}
\title{Large scale flow effects, energy transfer, and self-similarity 
       on turbulence}

\author{P.D. Mininni, A. Alexakis, and A. Pouquet}
\affiliation{NCAR, P.O. Box 3000, Boulder, Colorado 80307-3000, U.S.A.}
\date{\today}

\begin{abstract}
The effect of large scales on the statistics and dynamics of 
turbulent fluctuations is studied using data from high resolution 
direct numerical simulations. Three different kinds of forcing, and 
spatial resolutions ranging from $256^3$ to $1024^3$, are being used. 
The study is carried out by investigating the nonlinear triadic 
interactions in Fourier space, transfer functions, structure functions, 
and probability density functions. Our results show that the large scale 
flow plays an important role in the development and the statistical 
properties of the small scale turbulence. The role of helicity is 
also investigated. We discuss the link between these findings and 
intermittency, deviations from universality, and possible 
origins of the bottleneck effect. Finally, we briefly describe the 
consequences of our results for the subgrid modeling of turbulent flows.
\end{abstract}

\maketitle

\section{Introduction}

The nature of the interactions in a turbulent flow has been a long 
standing problem. Most investigations are focused on statistically 
homogeneous and isotropic turbulence. However, in most cases in nature 
this assumption is not necessarily true. Instead, one usually finds 
turbulence to be embedded in a large scale flow. In these cases, 
turbulence originates from some instability (e.g. convection) and is 
not uniformly distributed in space and time. To what extend the flow 
is approaching a statistically homogeneous and isotropic state as 
smaller and smaller scales are developed is still an open problem.

Due to insufficient computational resources (e.g in astrophysics, 
weather and climate prediction, for tokamaks, and in industrial 
applications) the computational efforts to study turbulent flows at 
these high Reynolds numbers are compelled to resort to modeling of the 
smaller scale turbulent fluctuations for which numerical resolution is 
not available. However, in order to accurately 
model the small scales a good understanding of the impact of the large 
scale flow in the small scales is needed, in addition to the knowledge 
of what (if any) is the feedback of the small scale turbulent flow 
on the large scale structures. This latter process is usually modeled 
by an effective viscosity, or other forms of subgrid modeling of the 
Reynolds stress tensor. These two types of interactions (from large to small 
and from small to large scales, or ``downscaling'' and ``upscaling'') 
are often considered to take place only through a local cascade of energy, 
although there is evidence that 
nonlocal interactions between widely separated scales are also relevant.

Studies of local and nonlocal interactions have been done previously 
in direct numerical simulations (DNS), although at moderate spatial 
resolutions and Reynolds numbers \cite{Domaradzki88,Domaradzki90,Kerr90,
Yeung91,Okhitani92,Zhou93,Zhou93b,Brasseur94,Yeung95,Zhou96,Kishida99}.
While some of these studies supported the existence of nonlocal 
interactions, it was argued by later studies that 
their presence was associated with moderate values of the 
Reynolds number in the simulations, or that it was linked to the precise 
definitions used for transfer functions and the interpretation of 
the results. In some cases \cite{Okhitani92,Zhou93,Zhou93b} large 
eddy simulations (LES) were also used to extend the range of Reynolds 
numbers studied, although if nonlocal interactions are present the impact 
of the subgrid model on the transfer is unclear and is a point of study 
in itself. Some of these studies 
also considered the effect of anisotropic or coherent large scale forcing 
(see e.g. Refs. \cite{Yeung91,Yeung95,Zhou96}). 
Evidence of nonlocal interactions between large and small scales has 
been found also in experiments \cite{Wiltse93,Wiltse98,Carlier01}. 
Observations of the persistence of anisotropy at small scales in 
experiments at large Reynolds numbers \cite{Sreenivasan97,Shen00}, in 
the atmosphere \cite{Stewart69}, and in numerical simulations 
(see e.g. Ref. \cite{Biferale01a} and references therein) also suggest 
the existence of a direct coupling between disparate scales. The presence 
of nonlocal interactions has also been associated in numerical studies 
with departures from universality and the development of intermittency 
\cite{Laval01}. Recently, a new study at high resolution 
\cite{Alexakis05b} presented a detailed analysis of nonlocal interactions 
using DNS for large Reynolds numbers, although for a particular and 
coherent forcing function.

In this work we examine three different kinds of flows in DNS in 
triple periodic boxes, generated by different body forces. We explore 
external forcings with and without helicity, and forcings with either infinite 
or short correlation times. The energy injection scale is varied, as 
well as the viscosity and spatial resolution of the runs, to explore a 
wide range of Reynolds numbers and configurations. In all cases, we 
observe the presence of nonlocal interactions coexisting with a local 
direct cascade of energy. The intensity of the nonlocal interactions 
depends on the transfer function studied. For a particular (highly 
symmetric) forcing, we also observe a correlation between regions of 
large scale shear and strong small scale gradients, supporting previous 
studies that linked intermittent effects with interactions with the 
large scale flow.

\section{Setup and theory}

\subsection{Equations, code, and simulations}

For an incompressible fluid with constant mass density, the Navier-Stokes 
equations are:
\begin{equation}
\partial_t {\bf u} + {\bf u}\cdot \nabla {\bf u} = - \nabla p 
    + \nu \nabla^2 {\bf u} +{\bf f} ,
\label{eq:momentum}
\end{equation}
\begin{equation}
\nabla \cdot {\bf u} =0, 
\label{eq:incompressible}
\end{equation}
where ${\bf u}$ is the velocity field, $p$ is the pressure divided by 
the mass density, and $\nu$ is the kinematic viscosity. Here, ${\bf f}$ 
is an external force that drives the turbulence. The mode with the largest 
wavevector in the Fourier transform of ${\bf f}$ is going to be denoted 
as $k_F$ and we are going to refer to $2 \pi k_F^{-1}$ as the forcing 
scale. We also define the viscous dissipation wavenumber as 
$k_\nu=(\epsilon/\nu^3)^{1/4}$, where $\epsilon$ is the energy injection 
rate (as a result, the Kolmogorov scale is $\eta = 2\pi/k_\nu$). A large 
separation between the two scales ($k_F^{-1} \gg  k_\nu^{-1}$) is required 
for the flow to reach a turbulent state.

In the absence of external forcing and viscosity, the Navier-Stokes 
equations in three dimensions have two ideal invariants: the energy
\begin{equation}
E = \frac{1}{2} \int{u^2 d {\bf x}^3} \, ,
\end{equation}
and the helicity
\begin{equation}
H = \frac{1}{2} \int{{\bf u} \cdot \nabla \times {\bf u} \, d{\bf x}^3} \, .
\end{equation}

All the results discussed in the following sections result from analysis 
of data from direct numerical simulations of the Navier-Stokes equations. 
We solve Eqs. (\ref{eq:momentum}) and (\ref{eq:incompressible}) using a 
parallel pseudospectral code in a three dimensional domain of size $2\pi$ 
with periodic boundary conditions \cite{Gomez05a,Gomez05b}. The pressure 
is obtained by taking the divergence of Eq. (\ref{eq:momentum}), using the 
incompressibility condition (\ref{eq:incompressible}), and solving the 
resulting Poisson equation. The equations are evolved in time using a 
second order Runge-Kutta method. The code uses the $2/3$-rule for 
dealiasing, and as a result the maximum wavenumber is $k_{max} = N/3$ 
where $N$ is the number of grid points in each direction. All simulations 
presented are well resolved, in the sense that the dissipation wavenumber 
$k_\nu$ is smaller than the maximum wavenumber $k_{max}$ at all times.

The Reynolds number is defined as $R_e = UL/\nu$, where $U$ is the 
r.m.s. velocity and $L$ is the integral lengthscale of the flow
\begin{equation}
L = 2\pi \frac{\int{E(k) k^{-1} dk}}{\int{E(k) dk}},
\label{eq:integral}
\end{equation}
where $E(k)$ is the energy spectrum. The large scale turnover time 
can then be defined as $T=U/L$. We can also introduce the Taylor 
based Reynolds number $R_\lambda = U\lambda/\nu$, where the Taylor 
lengthscale $\lambda$ is given by
\begin{equation}
\lambda = 2\pi \left(\frac{\int{E(k) dk}}{\int{E(k) k^2 dk}}\right)^{1/2}.
\label{eq:taylor}
\end{equation}

Several simulations were done with different resolutions (from $N=256$ to 
$1024$) and kinematic viscosities. Table \ref{table:runs} shows the 
parameters for all the runs. The rms velocity in the steady state of all 
the runs is $U\approx 1$. To asses the effect of different large scale 
stirring forces, we used three expressions for the external force 
${\bf f}$. The first expression corresponds to a Taylor-Green (TG) flow 
\cite{Taylor37}
{\setlength\arraycolsep{2pt}
\begin{eqnarray}
{\bf f}_{\rm TG} &=& f_0 \left[ \sin(k_F x) \cos(k_F y) 
     \cos(k_F z) \hat{x} - \right. {} \nonumber \\
&& {} \left. - \cos(k_F x) \sin(k_F y) 
     \cos(k_F z) \hat{y} \right] ,
\label{eq:TG}
\end{eqnarray}}
where $f_0$ is the force amplitude. This expression is not a solution of 
the Euler's equations, and as a result small scales are generated fast 
when the fluid is stirred with this forcing. The resulting flow has no net 
helicity, although regions with strong positive and negative helicity 
develop.

In order to study directly the effect of helicity and its transfer 
between different scales, 
we also did simulations using the Arn'old-Childress-Beltrami (ABC) forcing
{\setlength\arraycolsep{2pt}
\begin{eqnarray}
{\bf f}_{\rm ABC} &=& f_0 \left\{ \left[B \cos(k_F y) + 
    C \sin(k_F z) \right] \hat{x} + \right. {} \nonumber \\
&& {} + \left[A \sin(k_F x) + C \cos(k_F z) \right] \hat{y} + 
   {} \nonumber \\
&& {} + \left. \left[A \cos(k_F x) + B \sin(k_F y) \right] 
   \hat{z} \right\},
\label{eq:ABC}
\end{eqnarray}}
with $A=0.9$, $B=1$, and $C=1.1$. The ABC flow is an eigenfunction of 
the curl and an exact solution of the Euler equations. When the flow 
is forced using this expression, turbulence develops after an 
instability sets in \cite{Podvigina94}. Unlike TG forcing, ABC forcing 
injects net helicity into the flow.

The amplitude and phase of these two forcings is kept constant during the 
simulations, and as a result the external force has an infinite correlation 
time. It is a common practice in studies of isotropic and homogeneous 
turbulence to force in Fourier space injecting energy in all modes 
in a Fourier shell and changing the phase of each mode with a short 
correlation time. To compare with the results of TG and ABC forcing, we 
also implemented a random forcing
\begin{equation}
{\bf f}_{\rm RND} = f_0 \sum_{|{\bf k}|=k_F} i {\bf k} \times 
    \hat{\bf x} \, e^{i ({\bf k} \cdot {\bf x} + \phi_{\bf k})}
\label{eq:random}
\end{equation}
where the phase $\phi_{\bf k}$ of each mode with wavevector ${\bf k}$ was 
changed randomly with a correlation time $\tau_c$ that was taken to be
$\tau_c=0.1 T$.

\begin{table}
\caption{\label{table:runs}Parameters used in the simulations. $N$ is 
         the linear grid resolution, ${\bf f}$ the forcing [either Taylor 
         Green (TG), Beltrami (ABC) or random (RND)], $k_F$ 
         the forcing wavenumber, $\nu$ the kinematic viscosity, $R_e$ 
         the Reynolds number, and $R_\lambda$ the Taylor based Reynolds 
         number.}
\begin{ruledtabular}
\begin{tabular}{ccccccc}
Run & $N$  & ${\bf f}$ & $k_F$ &     $\nu$         & $R_e$ & $R_\lambda$ \\
\hline
I   & 256  &  TG       &   2   &$2\times 10^{-3}$  &  675  &     300     \\
II  & 512  &  TG       &   2   &$1.5\times 10^{-3}$&  875  &     350     \\
III & 1024 &  TG       &   2   &$3\times 10^{-4}$  & 3950  &     800     \\
IV  & 256  &  ABC      &  10   &$2.5\times 10^{-3}$&  275  &     230     \\
V   & 256  &  ABC      &   3   &$2\times 10^{-3}$  &  820  &     360     \\
VI  & 512  &  ABC      &   3   &$6.2\times 10^{-4}$& 2520  &     670     \\
VII & 1024 &  ABC      &   3   &$2.5\times 10^{-4}$& 6200  &    1100     \\
VIII& 256  &  RND      &   1   &$1.5\times 10^{-3}$& 2030  &     650     \\
\end{tabular}
\end{ruledtabular}
\end{table}

\subsection{Scale interactions and transfer}
To investigate the interactions between different scales we split the
velocity field into spherical shells in Fourier space of unit width,
\ie $\bv=\sum_K {\bf v}_K$ where ${\bf v}_K$ is a filtered velocity
field such that only the Fourier modes in the shell $K\le|k|<K+1$  
(from now on called shell $K$) are kept. 
From equation 
(\ref{eq:momentum}), the rate of energy transfer $T_3(K,P,Q)$ (a
third-order correlator) from energy in shell $Q$ to energy in shell $K$ 
due to the interaction with the velocity field in shell $P$ is defined 
as usual \cite{Kraichnan71,Lesieur,Alexakis05,Alexakis05b} as:
\begin{equation}
T_3(K,P,Q) = -\int \bv_K \cdot (\bv_P \cdot \nabla) \bv_Q d{\bf x}^3 \ .
\label{trans_eq3}
\end{equation}
Note that this term does not give information about the energy the
shell $P$ receives or gives to the shells $K$ and $Q$. 
The computation of $T_3$ for the three shells $K$, $P$, and $Q$ 
from 1 up to a wavenumber $K_{max}$ requires $\sim K_{max}^3 N^3$ 
operations and is therefore demanding in computer resources. For example, 
in the $1024^3$ simulations, to compute $T_3(K,P,Q)$ for all wavenumbers 
up to $K_{max}=80$, it takes as much computing time as 
evolving the hydrodynamic code for two turnover times.

If we sum over the middle wave number $P$ we obtain the total energy
transfer $T_2(K,Q)$ from shell $Q$ to shell $K$:
\begin{equation}
T_2(K,Q) = \sum_P T_3(K,P,Q) = -\int \bv_K \cdot (\bv \cdot \nabla)
\bv_Q d{\bf x}^3 \ .
\label{trans_eq2}
\end{equation}
Positive transfer implies that energy is transfered from shell $Q$ to $K$,
and negative from $K$ to $Q$; thus, both $T_3$ and $T_2$ are antisymmetric
in their $(K,Q)$ arguments (see Ref. \cite{Alexakis05}). $T_2(K,Q)$ gives 
information on the shell-to-shell energy transfer between $K$ and $Q$, 
but not about the amplitude of the triadic interactions 
themselves.

If we further sum over the wave number $Q$ we obtain the transfer
\begin{equation}
T_1(K) = \sum_Q T_2(K,Q) = -\int \bv_K \cdot (\bv \cdot \nabla) \bv d{\bf x}^3 
\label{trans_eq1}
\end{equation}
that expresses the rate the shell $K$ receives energy from the velocity 
field (all shells).

The energy flux is reobtained from these transfer functions as
\begin{equation}
\Pi(k) = -\sum_{K=0}^k T_1(K)=-\int \bv_K^< (\bv\cdot \nabla )\bv_K d{\bf x}^3
\label{eq:flux}
\end{equation}
where the notation
\begin{equation}
\bv_K^<=\sum_{K'=0}^K \bv_{K'} \,\,\, \mathrm{and} \,\,\, 
\bv_K^>=\sum_{K'=K+1}^\infty \bv_{K'} 
\end{equation}
is used (see Ref. \cite{Frisch}). 

We can further define the flux of energy at some given scale due to 
the interactions with some other scale as
\begin{equation}
\Pi_P(k) = -\sum_{K=0}^k \sum_Q T_3(K,P,Q)=
-\int \bv_K^< \cdot (\bv_P \cdot \nabla) \bv^> d{\bf x}^3 \,\, ,
\end{equation}
which expresses the flux of energy at the scale $K^{-1}$ due to the 
interactions with the scale $P^{-1}$
or in other words the energy flux at the scale $K^{-1}$ if the velocity
field was advected just by the velocity field at scale $P^{-1}$.
As will be discussed later, the 
question of locality depends on which of the different transfer 
functions or flux is under investigation.

It is worth noticing that through the study of the three functions 
$T_1$, $T_2$, $T_3$, and the fluxes $\Pi_P$, we can measure the nature 
and intensity of the interactions directly, 
and therefore we avoid 
introducing a non-locality parameter as was done e.g. in Refs. 
\cite{Zhou93,Zhou93b}.

The total energy balance equation for a given shell is written as
\begin{equation}
\partial_t E(K) = T_1(K)+ \nu {\mathcal D(K)} + {\mathcal F}(K)
\end{equation}
where we have also introduced the dissipation function
\begin{equation}
\nu {\mathcal D}(K) \equiv  \nu   \int |{\bf \nabla u}_K|^2  d{\bf x}^3 ,
\end{equation}
and the energy injection rate to the velocity field through the forcing term
\begin{equation}
{\mathcal F}(K) \equiv \int { \bf f} \cdot {\bf u}_K  \, d{\bf x}^3 \,.
\end{equation}

Finally, we will also investigate the transfer of helicity 
among different scales. Taking the inner product of the 
Navier-Stokes equation (\ref{eq:momentum}) with the vorticity at a scale $K^{-1}$
($\bw_K=\nabla \times \bu_K$) and adding the inner product of the velocity at 
the same scale $K^{-1}$ with the $curl$ of (\ref{eq:momentum}) 
and space averaging we obtain:
\begin{eqnarray}
\partial_tH(K)=&\sum_Q \int \bw_K \cdot (\bu \times \bw_Q) d{\bf x}^3+ 
    \nonumber \\
&\int \bw_K \cdot F d{\bf x}^3 +\nu \int \bw_K \cdot \nabla \times 
    \bw_K d{\bf x}^3.
\end{eqnarray}
Each term of the sum in the first term in the $r.h.s$ 
of the equation above can be written as:
\begin{equation}
T_H(K,Q) = \int \bw_K \cdot (\bu \times \bw_Q) d{\bf x}^3 \ .
\label{trans_eqh}
\end{equation}
Note that the anti-symmetric property $T_H(K,Q)=-T_H(Q,K)$ holds for
$T_H(K,Q)$ \ie the rate the shell $K$ is gaining helicity from the 
interaction with the field $\bw_Q$ and $\bu$ is equal to the rate the 
shell $Q$ is loosing helicity through the same interaction. This allows 
us to interpret $T_H(K,Q)$ as the transfer of helicity from the 
scale $Q^{-1}$ to the scale $K^{-1}$ (see \cite{Alexakis06} for the 
equivalent definition of the magnetic helicity transfer).

\section{\label{sec:runIII}Influence of the large scale flow on turbulence}

We discuss in this section properties of Run III. First we describe the 
geometry of the resulting flow and we present general statistical results, 
such as probability density functions and structure functions. Then, we 
analyze in detail the scale interactions using the formulation discussed 
in the previous section. Preliminary results for this flow were presented 
in \cite{Alexakis05b}. The results discussed here will be used as a 
reference to compare with the rest of the simulations in the following 
sections.

\subsection{Statistical properties, structure functions, and scaling exponents}

\begin{figure}
\centerline{\includegraphics[width=8.3cm]{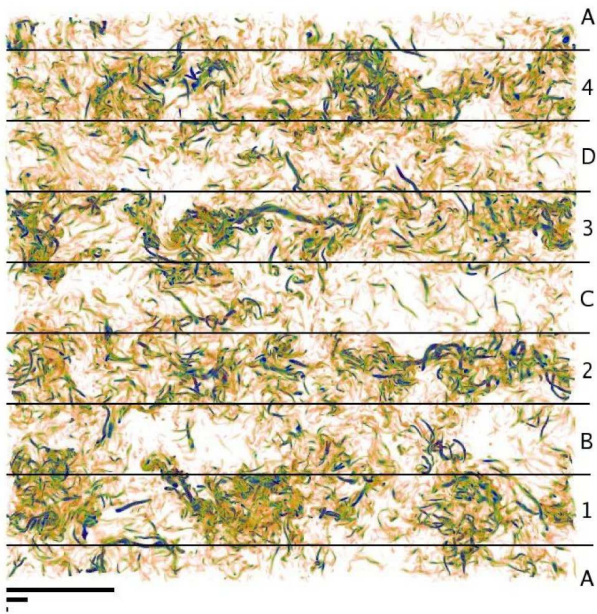}}
\caption{(Color online) Rendering of enstrophy density in a 
         $1024 \times 64 \times 1024$ 
         slice of Run III. Bands of strong (1,2,3,4) and weak (A,B,C,D) 
         shear in the external Taylor-Green force are indicated. The bars 
         at the bottom show respectively the integral, Taylor, and 
         dissipation scales.}
\label{fig:render}
\end{figure}

Run III is a $1024^3$ simulation using TG forcing. After reaching a 
turbulent steady state, the simulation was continued for 10 turnover 
times. The expression of the forcing has several spatial symmetries, and 
some of these symmetries are recovered in the flow in a statistical 
sense. Figure \ref{fig:render} shows a rendering of enstrophy density in a 
thin slice of $1024 \times 64 \times 1024$ grid points. Thin and elongated 
structures (vortex tubes) can be identified. As observed in previous 
studies, these structures are characterized by one large lengthscale 
in one direction (ranging from the Taylor to the integral scale) and 
a short lengthscale (close to the viscous dissipation scale) in the other 
two directions. 

\begin{figure}
\centerline{\includegraphics[width=7cm]{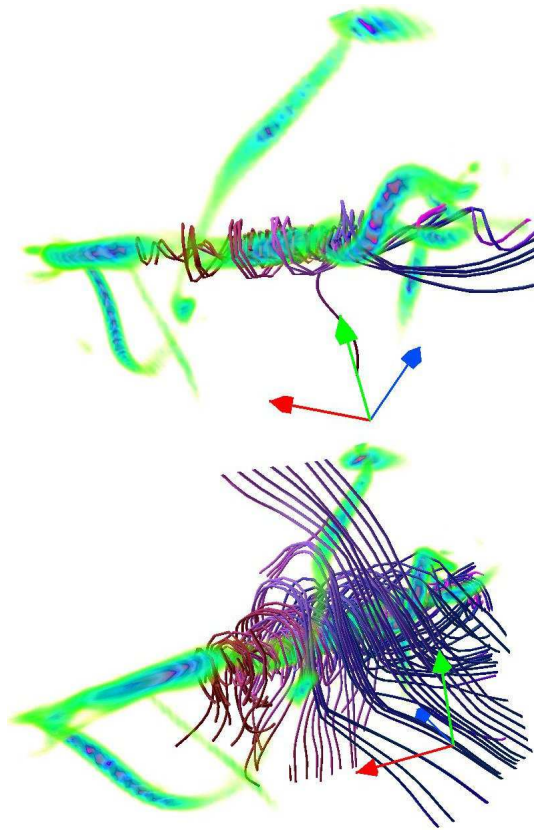}}
\caption{(Color online) Zoom on a region with large enstrophy density 
         showing field lines inside 
         (above) and in its surrounding (below) in Run III. Velocity field 
         lines are helical inside and in the vicinity of the tube. Note the 
         merging of two tubes in the south west corner. The side of the region 
         shown is approximately one tenth of the side of the total 
         domain.}
\label{fig:tube}
\end{figure}

Since $k_F=2$, there are four planes in the $z$ direction where the 
external force is zero (see Eq. \ref{eq:TG}) and the shear is maximum. 
In the steady state turbulent flow, these planes can be easily recognized 
since four bands where the vorticity is stronger are formed around them. 
Regions with less (or weaker) vortex tubes separate these planes, 
regions centered around the planes where the large scale shear has 
a minimum. Together, these two sets of regions 
form a large scale pattern that is observed to persist for long times, 
giving four ``quiet'' stripes and four stripes of ``stronger" turbulence 
(the boundary between these regions is not well defined and fluctuates 
in space and time, see Fig. \ref{fig:render}). Considering the persistence 
of these statistical 
symmetries of the flow, the velocity increments for this Run will be 
computed for displacements {\it only} in the $x$-$y$ plane.

\begin{figure}
\centerline{\includegraphics[width=8.3cm]{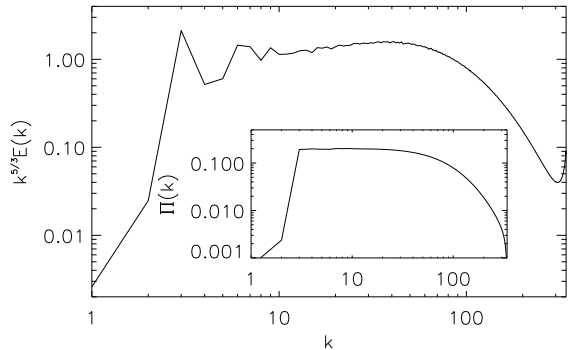}}
\caption{Energy spectrum compensated by $k^{-5/3}$ for Run III. The 
         inset shows the energy flux.}
\label{fig:spectrum_runIII}
\end{figure}

When individual vortex tubes are studied, it is seen that the 
flow inside and surrounding the vortex tube is helical, as noted before by several authors
\cite{Tsinober83,Moffatt85,Moffatt86,Levich87,Farge01}. Figure 
\ref{fig:tube} shows a rendering of a small region in the domain 
(approximately one tenth of the box). A region of large enstrophy 
density is indicated by dark (blue) colors, and field lines inside 
and in the vicinity of the enstrophy containing region are shown. 
In this region and its surroundings, field lines are helical, while 
the flow far from the region is not. This feature has been verified 
for a large set of vortex tubes in the complete domain. It has been 
claimed in the past \cite{Moffatt92,Tsinober} that the development of these 
helical structures in a turbulent flow can lead to the depletion of 
nonlinearity and a quenching of local interactions. We will come back 
to this issue in the following sections.

\begin{figure}
\centerline{\includegraphics[width=8.3cm]{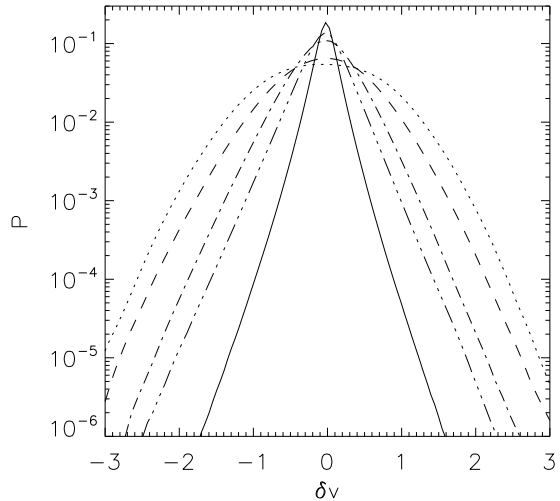}}
\caption{Probability density functions of transverse velocity increments 
         for Run III, for increments $l=20 \eta$ ($\cdots$), $l=10 \eta$ 
         ($- - -$), $l=4 \eta$ ($-\cdot-$), $l=2 \eta$ ($-\cdots-$), and 
         $l=\eta$ (solid line), where $\eta$ is the Kolmogorov dissipation 
         scale.}
\label{fig:pdf_runIII}
\end{figure}

In spite of the presence of the large scale pattern, many features often 
associated with isotropic and homogeneous turbulence can be observed 
in this simulation. Figure \ref{fig:spectrum_runIII} shows the 
angle-averaged Fourier energy spectrum and energy flux. An inertial 
range with constant energy flux is observed, together with a 
Kolmogorov-like scaling and a bottleneck as the dissipative range is 
reached. When probability density functions (pdfs) of transverse 
velocity increments
\begin{equation}
\delta v_\perp ({\bf x},l) = \hat{\bf r} \times \left[ {\bf v}({\bf x} + 
    l\hat{\bf r}) - {\bf v}({\bf x}) \right],
\label{eq:incrementperp}
\end{equation}
are computed in the whole domain (Fig. \ref{fig:pdf_runIII}), we observe 
distributions close to Gaussian for large increments, and the development 
of exponential and stretched exponential tails as the increment $l$ is 
decreased.

\begin{figure}
\centerline{\includegraphics[width=8.3cm]{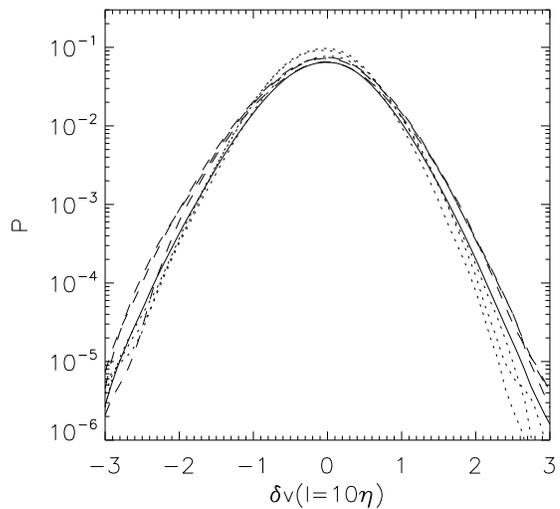}}
\caption{Probability density functions of transverse velocity increments 
         with $l=10 \eta$ for Run III. The solid line corresponds to the 
         whole domain, dashed lines to regions 1 to 4 (strong shear), 
         and dotted lines to regions A to D (weak shear; see Fig. 
         \ref{fig:render}). Notice the weak but systematic differences:
         the four dashed lines (some of them overlapping) have slower decaying 
         tails and the dotted lines have faster decaying tails.}
\label{fig:pdf10eta}
\end{figure}

\begin{figure}
\centerline{\includegraphics[width=8.3cm]{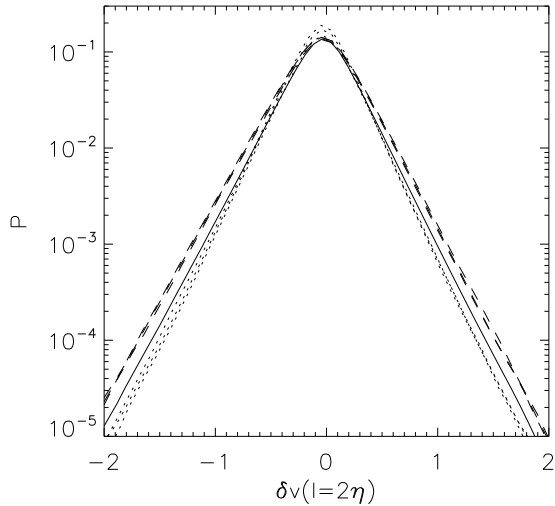}}
\caption{Probability density functions of transverse velocity increments 
         with $l=2 \eta$ for Run III. Same labels as in Fig. 
         \ref{fig:pdf10eta}. Note that the differences between quiet 
         and strong regions appear even more systematic at this scale than in Fig. 
         \ref{fig:pdf10eta}.}
\label{fig:pdf2eta}
\end{figure}

Figures \ref{fig:pdf10eta} and \ref{fig:pdf2eta} show pdfs of transverse 
velocity increments with $l=10 \eta$ and $2 \eta$ respectively, but 
discriminating between bands of strong and weak shear (as indicated in Fig. 
\ref{fig:render}). Bands with strong shear (1,2,3, and 4) have 
slightly but systematically stronger 
tails (i.e. a larger probability of strong gradients) than regions A,B,C, 
and D. Note this behavior is true for each individual region, and the 
difference increases as the increment $l$ is decreased. This confirms 
statistically what can be inferred from Fig. \ref{fig:render}: regions 
of strong shear have a larger density of vortex tubes and stronger 
gradients.

\begin{figure}
\centerline{\includegraphics[width=8.3cm]{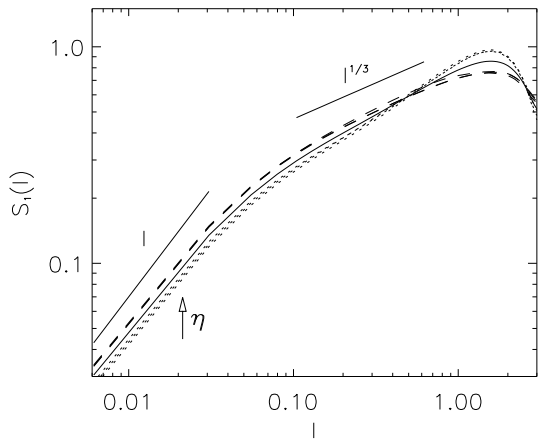}}
\caption{First order structure function $S_1(l)$ for Run III. The solid 
         line corresponds to the whole domain, dashed lines to regions 
         1,2,3, and 4, and dotted lines to regions A,B,C, and D. The 
         Kolmogorov scaling and the linear slope corresponding 
         to a smooth flow are shown as a reference. The arrow indicates 
         the Kolmogorov scale $\eta$. Notice again the systematic 
         differences between the two regimes identified in Fig. 1.}
\label{fig:S1}
\end{figure}

\begin{figure}
\centerline{\includegraphics[width=8.3cm]{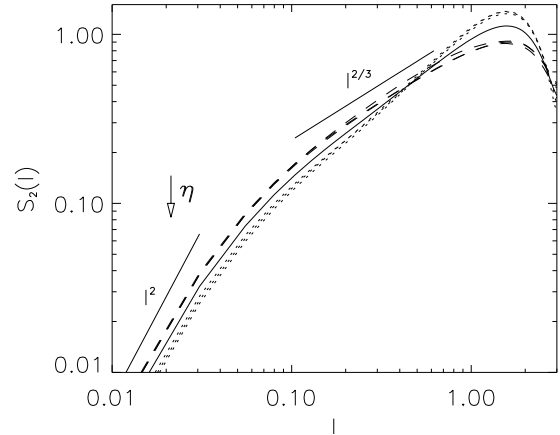}}
\caption{Second order structure function $S_2(l)$ for Run III. Labels  
         are as in Fig. \ref{fig:S1}.}
\label{fig:S2}
\end{figure}

We can also compute longitudinal velocity increments
\begin{equation}
\delta v_\parallel ({\bf x},l) = \hat{\bf r} \cdot \left[ {\bf v}({\bf x} + 
    l \hat{\bf r}) - {\bf v}({\bf x}) \right] .
\label{eq:incrementpar}
\end{equation}
Figures \ref{fig:S1} and \ref{fig:S2} show respectively the second and 
third order longitudinal structure functions, where the structure function 
of order $p$ is defined as
\begin{equation}
S_p (l) = \left< \delta v_\parallel^p ({\bf x},l) \right> .
\label{eq:structure}
\end{equation}
At scales smaller than the dissipation scale the field is smooth and 
the structure function of order $p$ scales as $l^p$. Kolmogorov's 1941 
theory of turbulence predicts in the inertial range a scaling 
$S_p (l) \sim l^{p/3}$, although corrections due to intermittency are 
known and in general $S_p (l) \sim l^{\zeta_p}$, where $\zeta_p\not= p/3$ are 
the scaling exponents.

As with pdfs of velocity increments, a clear trend separating regions of 
strong and weak shear is observed. In the range of scales corresponding to 
the inertial range, the four regions with strong shear (regions 1,2,3 and 4) 
show a larger slope than the four regions with weak shear. The slope of the 
structure function computed in the whole box lies between these two 
values. Considering the results from both pdfs and structure functions, 
a correlation is observed between small scale gradients and large scale 
shear. Note that a correlation between stronger tails in the pdfs 
of velocity increments and vortex tubes has already been observed in 
\cite{Levi01,Farge01}. Here, a correlation between these quantities with 
large scale shear is further observed, in agreement with Ref. \cite{Laval01} 
that suggests intermittency is related with interactions with the large 
scale flow.

\subsection{Interactions, energy transfer, and flux}

\begin{figure}
\centerline{\includegraphics[width=8.3cm]{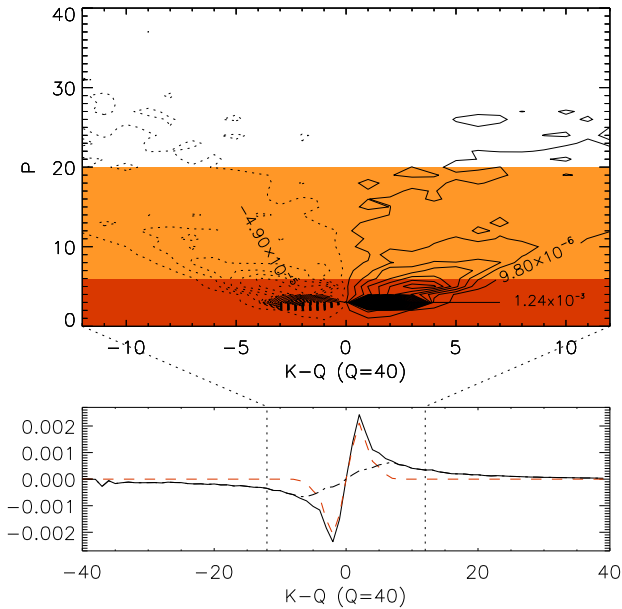}}
\caption{(Color online) Above: amplitude of triadic interactions 
         as given by $T_3(K,P,Q)$ 
         at $Q=40$ for Run III. Solid lines correspond to positive contour 
         levels, and dotted lines to negative values. The dark (red) shaded  
         region indicates shells in $P$ close to the forcing shell, while 
         the light (orange) shaded region indicates shells in $P$ outside 
         the local octave. Below: total shell-to-shell energy transfer 
         $T_2(K,Q)$, transfer $T_2^{\textrm LS}(K,Q)$ due to interactions 
         with $P$ close to the forcing shell (dashed line), and difference 
         between these two transfers (dash-dotted line).}
\label{fig:triads}
\end{figure}

Figure \ref{fig:triads} shows the functions $T_3(K,P,Q)$ and $T_2(K,Q)$
evaluated at $Q=40$ in the turbulent steady state of Run III. When triadic 
interactions between shells are studied, the strongest interactions are
with the shell $P=3$, where the large scale forcing is. Local 
interactions between Fourier shells ($K \sim P \sim Q$) are two orders 
of magnitude smaller than nonlocal interactions with $P \sim k_F$. 
As a result, when $T_3(K,P,Q)$ is summed over all $P$ to obtain the 
shell-to-shell energy transfer $T_2(K,Q)$, the energy is observed to 
be transfered to small scales locally between shells, but with a step 
proportional to $k_F$. In $T_2(K,Q)$, the negative peak at 
$K \sim Q-k_F$ means energy is transfered from this shell to the shell 
$Q$, while the positive peak at $K \sim Q+k_F$ indicates energy is 
transfered to this shell from the shell $Q$.

The function $T_3(K,P,Q)$ was studied in the steady state of this run 
also for $Q=10$ and $Q=20$, obtaining the same quantitative results: a 
dominance (when compared in amplitude) of triadic interactions with the 
large scale flow over local triadic interactions. 
The $T_2(K,Q)$ function for all values of $Q$ between $10$ and $80$ also 
peaks at $K\approx Q \pm k_F$ (see Ref. \cite{Alexakis05b}). These peaks 
at fixed values of $K-Q$ in $T_2(K,Q)$ are just the signature of the 
strong triadic interactions with the large scale forcing. If we compute 
the shell-to-shell transfer between shells $K$ and $Q$ due only to 
interactions with the large scale flow 
\begin{equation}
T_2^{\textrm LS} (K,Q) = \sum_{P=0}^6 T_3(K,P,Q) \, ,
\label{eq:T2LS}
\end{equation}
we obtain most of the shell-to-shell transfer (see Fig. \ref{fig:triads}). 
The remaining transfer $T_2(K,Q)-T_2^{\textrm LS}(K,Q)$ still peaks at 
larger wavenumbers, although it still does not peak at $K \sim Q = 40$. 
This is due to nonlocal interactions outside the octave 
band not related with the large scale forcing (light shaded region in 
Fig. \ref{fig:triads}). Note that the definition of the range summed over 
$P$ in Eq. (\ref{eq:T2LS}), or of octave bands (here defined as 
$Q/2 \lesssim K \lesssim 2Q$) is somewhat arbitrary.

\begin{figure}
\centerline{\includegraphics[width=8.3cm]{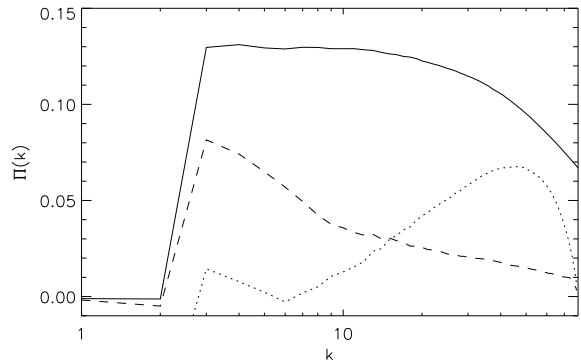}}
\caption{Total energy flux $\Pi(k)$ in Run III (solid line), flux 
         $\Pi^{\textrm LS}(k)$ due to interactions with the large scale 
         flow (dashed line), and nonlocal flux $\Pi^{\textrm NL}(k)$.}
\label{fig:fluxes_runIII}
\end{figure}

As previously mentioned, while individual triadic interactions as described 
by $T_3(K,P,Q)$ are dominantly nonlocal, the shell-to-shell transfer 
$T_2(K,P,Q)$ describes a local transfer of energy although through 
interactions with the large scale flow. Clearly when we sum over shells, 
the larger number of modes in the small scales start dominating over the 
large scale modes. Summing further over $K$ and $Q$, from Eq. 
(\ref{eq:flux}) we can obtain the total energy flux. In analogy with Eq. 
(\ref{eq:T2LS}) we can also define the flux due to interactions with the 
large scale flow (dark shaded region in Fig. \ref{fig:triads})
\begin{equation}
\Pi^{\textrm LS}(k) = \sum_{P=0}^{6} \Pi_P(k) \, ,
\end{equation}
and the nonlocal flux due to interactions outside an octave band (light 
shaded regions in Fig. \ref{fig:triads})
\begin{equation}
\Pi^{\textrm NL}(k) = \sum_{P=7}^{k/2} \Pi_P(k) \, ,
\end{equation}
where $\Pi_P(k)$ is defined in Eq. (15).

Figure \ref{fig:fluxes_runIII} shows these three fluxes in the steady 
state of Run III. The large scale flow is only responsible for a small 
(but not insignificant) fraction of the total energy flux 
($\Pi^{\textrm LS} \simeq 0.2 \Pi$).
Furthermore although the amplitude of the nonlocal flux 
$\Pi^{\textrm NL}(k)$ depends on the definition of an octave band, it is 
remarkable that it peaks at wavenumbers close to the peak of the 
bottleneck in the energy spectrum (see Fig. \ref{fig:spectrum_runIII}). 
This gives a direct confirmation that the bottleneck is due to the 
depletion of the energy transfer due to local interactions with 
$K \sim P \sim Q$. These interactions are inhibited because of the 
presence of a numerical and/or viscous cutoff in wavenumber, and 
nonlocal triads become dominant 
\cite{Herring82,Falkovich94,Lohse95,Martinez97}. Indeed, in Ref. 
\cite{Herring82} it was shown using the eddy-damped quasinormal Markovian 
(EDQNM) approximation that the bottleneck disappears if nonlocal 
interactions are excluded from the computation. Note that the spurious 
drop of the fluxes $\Pi^{\textrm LS}$ and $\Pi^{\textrm NL}$ at $k=80$ 
in Fig. \ref{fig:fluxes_runIII} is because the functions $T_3(Q,P,Q)$ 
and $T_2(K,Q)$ were only computed up to this wavenumber.

Therefore there seems to be a hierarchy concerning the importance of 
nonlocality when one investigates the transfer functions.
At the most basic level when one investigates the transfer function $T_3$
the interactions with the large scale flow are the strongest ones. 
When averaged over the middle wave number the resulting energy transfer $T_2$ 
becomes local but not self-similar with a maximum at $K-Q\sim k_F$. When 
averaged further to obtain the energy flux, the non local interactions with 
the large scale flow are only responsible for  20\% of the flux. 

\section{\label{sec:Reynolds}Scaling with Reynolds}

In this section we discuss results from runs I, II, and III. 
The three runs are forced using Eq. (\ref{eq:TG}), and the only parameters 
changed between the runs are the spatial resolution and the kinematic 
viscosity. This investigation allows us to study how the results about 
the non-local interactions change as we increase the Reynolds number.
Qualitatively the functions $T_3(K,P,Q)$ and $T_2(K,Q)$ are similar for 
all three runs: the strongest interactions [$T_3(K,P,Q)$] are with
the large scale flow and the energy transfer [$T_2(K,Q)$] is local with
the two peaks at $K-Q \sim k_F$.  
For this reason we focus here on the study of the local 
and nonlocal flux of energy as the Reynolds number is changed.

\begin{figure}
\centerline{\includegraphics[width=8.3cm]{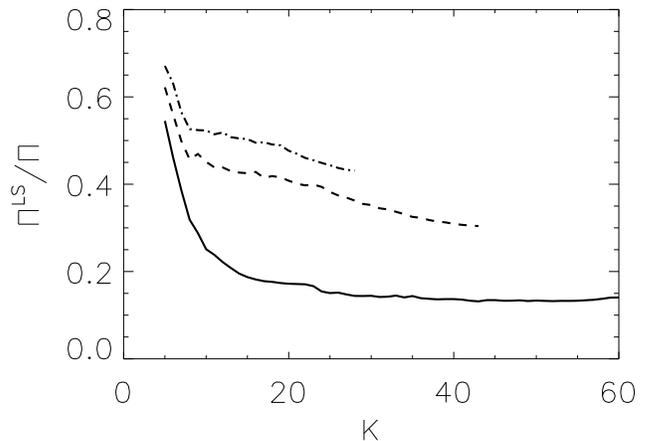}}
\caption{Scale variation of the flux ratio $\Pi^{\textrm LS}/\Pi$ for 
         increasing Reynolds numbers; Runs I (dash-dot line), 
         II (dash line), and III (solid line). Note the presence of a wavenumber range 
         where $\Pi^{\textrm LS}/\Pi$ becomes constant only for the 
         highest resolution run.}
\label{fig:fluxratio}
\end{figure}

In Figure \ref{fig:fluxratio} we show the ratio of the flux 
$\Pi^{\textrm LS}(k)$ to the total flux $\Pi(k)$. As discussed in the 
previous section, the flux due to interactions with the large flow gets 
diminished as we sum over more and more modes. As a result, in the 
inertial range of Run III interactions with the large scale flow are 
only responsible for $\sim 20\%$ of the energy flux. In the smaller 
Reynolds number runs (I and II), the amount of flux due to interactions 
with the large scale flow is larger. This implies that as we increase 
the Reynolds number the fraction of the energy flux in the inertial 
range due to interactions with the large scale flow decreases. 
However, note that in Run III a region in the inertial range where 
the ratio $\Pi^{\textrm LS}/\Pi$ is approximately constant is observed. 
This region with constant ratio $\Pi^{\textrm LS}/\Pi$ is not present 
in the simulations with lower Reynolds number. 
Therefore DNS with even higher Reynolds number than what is accomplished 
here are needed
to determine how does this ratio scale with the Reynolds number.

\section{Effects of different forcing functions}

\subsection{Forcing expression and correlation time}

\begin{figure}
\centerline{\includegraphics[width=8.3cm]{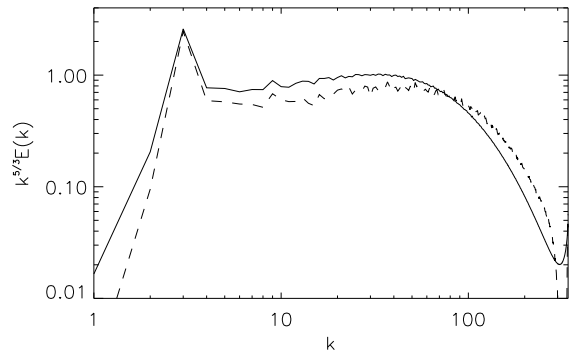}}
\caption{Compensated energy spectrum $k^{5/3} E(k)$ (solid line) and 
         helicity spectrum $k^{5/3} H(k)/k_F$ (dashed line) in the 
         steady state of Run VII, where $k_F$ is the forcing wavenumber.}
\label{fig:spectrum_runVII}
\end{figure}

Nonlocal interactions discussed in the previous section were observed 
for runs with large scale non-helical forcing with infinite correlation 
time. It is of interest to know how much of these results translate to 
other kind of forcings. In this section we compare results for runs 
III, VII and VIII. Run VII is a $1024^3$ simulation with constant helical 
forcing, and as runs I-III, it also displays a well defined large scale 
flow (for a description of the ABC flow, see e.g. Ref. \cite{Childress}). 
In Run VIII the phases of the forcing function are changed randomly with 
a short correlation time. Although the resolution in this run is smaller 
($256^3$), we will focus here on the effect of the forcing, since a 
systematic study of changing the Reynolds number for fixed forcing 
was presented in the previous section.

Figure \ref{fig:spectrum_runVII} shows the energy and helicity 
spectra compensated by a Kolmogorov law in the turbulent steady state of Run VII. 
As proposed in Ref. \cite{Brissaud73} and
observed in several simulations
\cite{Andre77,Borue97,Chen03,Chen03b,Gomez04} the spectrum of helicity 
follows a Kolmogorov law and is proportional to $k_F E(k)$. The transfer 
of helicity will be discussed in Section \ref{sec:helicity}.

\begin{figure}
\centerline{\includegraphics[width=8.3cm]{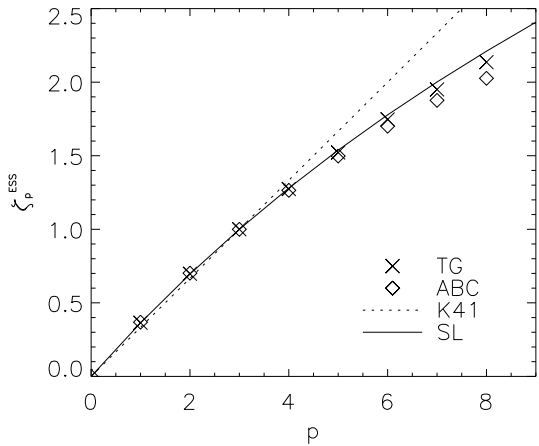}}
\caption{Scaling exponents of velocity structure functions $\zeta_p$ 
         as a function of order $p$ 
         in Runs III (TG forcing) and VII (ABC forcing). The Kolmogorov 
         K41 scaling \cite{Kolmogorov41}  is indicated, as well as the She-Leveque (SL) 
         prediction \cite{She94}. Note the difference between TG and ABC.} 
\label{fig:scaling}
\end{figure}

Figure \ref{fig:scaling} 
shows the scaling exponents $\zeta_p$ of the longitudinal structure 
functions, computed for runs III and VII. The extended self-similarity 
(ESS) hypothesis \cite{Benzi93,Benzi93b} was used to compute the anomalous 
exponents, which show similar behavior for both runs. However, 
the resulting exponents for $p>3$ from Run VII (ABC forcing) are slightly 
smaller than the exponents from Run III (TG forcing); for example,
 in Run III $\zeta_4 = 1.2737 \pm 0.0005$ and 
$\zeta_8 = 2.136 \pm 0.006$, while in Run VII $\zeta_4 = 1.2647 \pm 0.0005$ 
and $\zeta_8 = 2.026 \pm 0.008$. 
Note that the difference 
between the scaling exponents for TG and ABC at order $p=4$, though small, is
nevertheless more than order of magnitude larger than the error arising from
measuring them by virtue of the ESS hypothesis which leads to negligible 
errors.

It should be noted that it is unclear whether 
this observed difference is due to 
problems linked with using the ESS methodology itself, or if it indicates
a departure from universality between 
these two flows because of loss of homogeneity, as exemplified in Fig. 1.
Since the two forcing functions studied here are both to some degree
anisotropic and inhomogeneous, subleading contributions due to departures 
from full symmetry could be responsible for
this discrepancy. In this context, a decomposition of the structure 
functions into their isotropic and 
anisotropic components \cite{Biferale01a,Biferale01b,Biferale02} could 
help to asses the degree of universality of each component.
These points will necessitate further study and better resolved flows.

\begin{figure}
\centerline{\includegraphics[width=8.3cm]{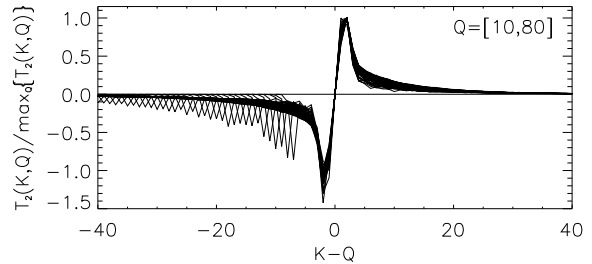}}
\caption{Shell-to-shell energy transfer function $T_2(K,Q)$ for Run VII. 
         The superimposed curves correspond to different values of $Q$, 
         with $Q\in [10,80]$.}
\label{fig:T2_runVII}
\end{figure}

The shell-to-shell energy transfer $T_2(K,Q)$ for Run VII is shown in Fig. 
\ref{fig:T2_runVII}. As for non-helical forcing (see Sec. \ref{sec:runIII}
and Ref. \cite{Alexakis05}) 
the shell-to-shell transfer function peaks at $K \approx Q \pm k_F$, 
indicating the transfer of energy is local but mediated by nonlocal 
interactions with the large scales. If we sum over $K$ and $Q$ to 
obtain the energy flux, we reobtain the results discussed in Sec. 
\ref{sec:Reynolds} for Run III: most of the energy flux is local, 
although in the inertial range $\sim 20\%$ of the total flux is due 
to interactions with the large scale forcing.

The transfer function $T_2(K,Q)$ was also computed in the turbulent 
steady state of Run VIII (see Fig. \ref{fig:T2_scale}.a). In this run 
the phases of the external force are changed randomly with a short 
correlation time. It is noteworthy that even in this case with 
isotropic and random forcing, evidence is found of nonlocal 
interactions with the large forcing scale: the $T_2(K,Q)$ 
function peaks at $K \approx Q \pm k_F$ for all wavenumbers $Q$ studied 
(note that in this run $k_F=1$). Fig. \ref{fig:T2_scale} is discussed further below.

\subsection{Forcing scale}

\begin{figure}
\centerline{\includegraphics[width=8.3cm]{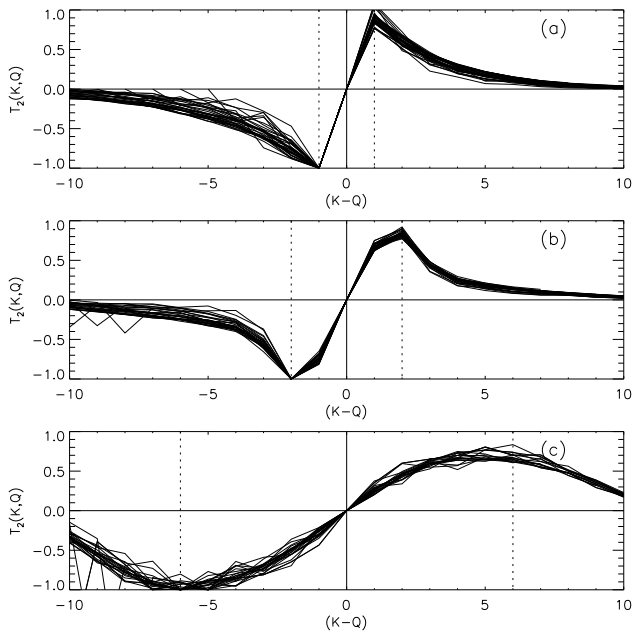}}
\caption{Shell-to-shell energy transfer function $T_2(K,Q)$ for runs with 
         different forcing wavenumbers $k_F$ as indicated by the vertical 
         lines; Run VIII with $k_F=1$ (a), Run V with $k_F=3$ (b), and 
         Run IV with $k_F=10$ (c). In each panel, the several curves 
         correspond to different values of $Q$ in the inertial and 
         dissipative ranges. Note that for each run, the peak of energy 
         transfer is centered close to $k_F$. }
\label{fig:T2_scale}
\end{figure}

Through all this work we have shown transfer functions indicating that the 
energy is locally transfered between scales but with a fixed step that 
we associated with the forcing scale. In this subsection we compare 
the transfer function $T_2(K,Q)$ in simulations forced at different 
wavenumbers.

Figure \ref{fig:T2_scale} shows $T_2(K,Q)$ for several 
values of $Q$ computed in the steady state of runs VIII ($k_F=1$), 
IV ($k_F=2$), and V ($k_F=10$). A clear correlation is observed 
between the position of the peaks ($|K-Q|$) in the shell-to-shell 
transfer function, and the forcing wavenumber $k_F$. In the last 
case, the peaks at $|K-Q|$ slightly smaller than $k_F$ could be 
associated with low Reynolds number effects.

\section{\label{sec:helicity}Helicity transfer}

The transfer of helicity is readily studied in Run VII, since the 
external forcing injects maximum helicity and all scales are dominated 
by the same sign of helicity. We will make no attempt here to separate 
the different signs of helicity in the simulations (see e.g. Refs. 
\cite{Waleffe91,Chen03,Chen03b}).

\begin{figure}
\centerline{\includegraphics[width=8.3cm]{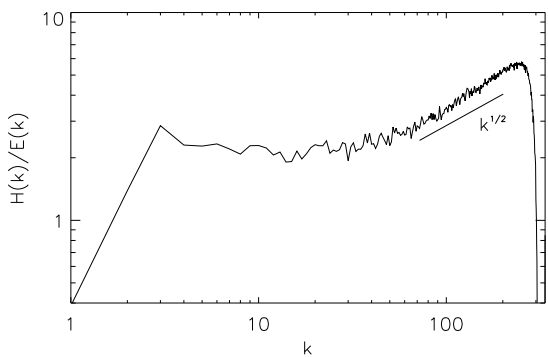}}
\caption{The spectrum of relative helicity $H(k)/kE(k)$
         compensated by $k^{-1}$ for run VII. Note that the region where the 
         spectrum is flat indicates a power law of $k^{-1}$ in the relative 
         helicity. At large wavenumber, a $k^{1/2}$ slope is shown only
         as a reference. The increase at large $k$ is indicative of an 
         excess of relative helicity in the small scales.}
\label{fig:relative}
\end{figure}

Before discussing the transfer of helicity, it is of interest to study 
spectral properties of helical flows. As previously discussed, the spectrum 
of helicity follows an approximate $k^{-5/3}$ law (see Fig. 
\ref{fig:spectrum_runVII}). As a result, the relative helicity $H(k)/E(k)k$ 
follows in the inertial range a $k^{-1}$ slope 
\cite{Brissaud73,Andre77,Lesieur} (i.e. small scales are less helical 
than large scales). 
It was predicted in Refs. \cite{Ditlevsen01,Ditlevsen01b} 
that the dissipation scale of helicity should be larger than the energy 
dissipation scale, giving as a result a drop in the spectrum of relative 
helicity faster than $k^{-1}$ for small scales. This argument would be 
in agreement with the idea that small scales slowly recover the mirror 
symmetry broken by the injection of helicity in the large scales. However, 
previous simulations \cite{Gomez04} and this high resolution run both suggest that
there is an excess of relative helicity in the small scales, when compared 
with the $k^{-1}$ drop. Note that this slower than predicted recovery of 
symmetries in the small scales is also in agreement with the slower than 
expected recovery of isotropy observed in experiments 
\cite{Sreenivasan97,Shen00}.

Figure \ref{fig:relative} shows $H(k)/E(k)$, i.e. the spectrum of relative 
helicity $R(k)=H(k)/kE(k)$ compensated by $k^{-1}$. 
A scaling of $k^{-1}$ for the relative helicity $R(k)$ thus corresponds 
to a flat spectrum in Fig. 16, 
as observed through the inertial range up to $k\sim 20$. However, at small 
scales the compensated spectrum of relative helicity grows, 
possibly as $k^{1/2}$ or steeper, indicating that
the spectrum of helicity $H(k)$ at small scales is dropping slower than 
the spectrum of energy $E(k)$ (see also Fig. \ref{fig:spectrum_runVII}). 
This can be associated with the presence of vortex tubes, that are 
known to be helical (see Fig. \ref{fig:tube}). 

\begin{figure}
\centerline{\includegraphics[width=8.3cm]{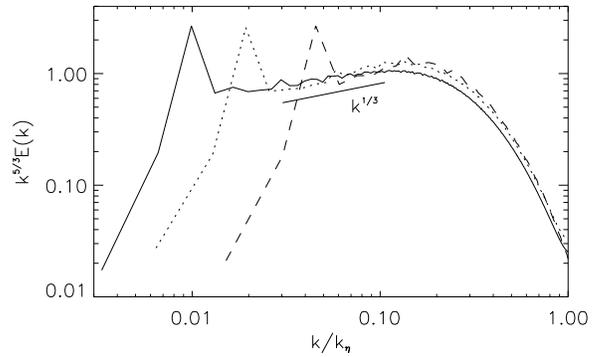}}
\caption{Energy spectrum compensated by $k^{-5/3}$ for Runs V (dashed 
         line), VI (dotted line), and VII (solid line). All runs are for 
         ABC forcing at different Reynolds numbers. Note the wavenumber 
         on the x-axis has been divided by the dissipation wavenumber to 
         make all dissipation ranges coincide. A slope of $1/3$ close to 
         the bottleneck (corresponding to a power law of $k^{-4/3}$ in 
         the energy spectrum) is shown as a reference.}
\label{fig:bottleneck}
\end{figure}

The effect of helicity, and indirectly the effect of nonlocal interactions 
which can be modeled for example
through the introduction of a second time-scale in the problem in order to 
distinguish between the eddy turn-over time and a helical characteristic 
time, has also been used to explain the development of the bottleneck 
effect that occurs at the onset of the dissipative range, see e.g. 
\cite{Kurien04}. Ref. \cite{Kurien04} predicts a $k^{-4/3}$ energy 
spectrum for the bottleneck, that we found compatible with our spectra 
for runs V-VII (Fig. \ref{fig:bottleneck}). We also observe a $k^{-4/3}$ 
range close to the bottleneck in the Taylor-Green flow that has no net 
helicity. Note that the argument in Ref. \cite{Kurien04} is based 
on the presence of non zero helicity locally. The transition from the 
inertial range to the bottleneck seems to be dependent on the Reynolds 
number. It is worth noting that while at resolutions of $256^3$ mostly 
a bottleneck is observed, in the $1024^3$ runs a short Kolmogorov-like 
scaling is found before the bottleneck takes place. The origin of the 
bottleneck based on the dominance of nonlocal interactions in the 
dissipative range has also been investigated 
\cite{Herring82,Falkovich94,Lohse95,Martinez97} as we discussed in 
Sec. \ref{sec:runIII}, and is independent of the presence of helicity.
These arguments are not mutually exclusive, as local in space generation 
of helicity in the small scales can also quench local interactions between 
eddies of comparable sizes. 
However, the prediction in Ref. \cite{Falkovich94} for the spectral 
shape of the bottleneck is in disagreement with the spectra obtained 
in all simulations here. These points deserve further study at higher 
resolution if one is to be able to distinguish between the different 
ranges that may be occurring.

\begin{figure}
\centerline{\includegraphics[width=8.3cm]{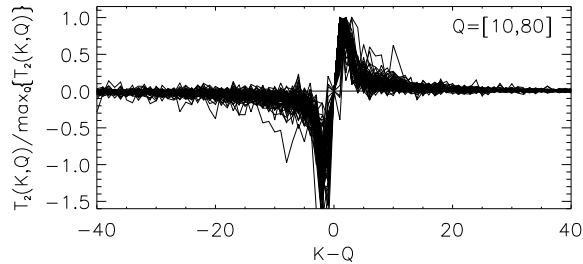}}
\caption{Shell-to-shell helicity transfer $T_H(K,Q)$ in Run VII, for 
         all values of $Q$ from $10$ to $80$.}
\label{fig:T2H}
\end{figure}

Finally, we discuss the shell-to-shell helicity transfer $T_H(K,Q)$ 
in Run VII (Fig. \ref{fig:T2H}). As for the transfer of energy, 
the helical transfer function peaks at $K \approx Q \pm k_F$, although this 
transfer is noisier. The helicity cascades directly to smaller scales 
as the energy, confirming previous studies 
\cite{Borue97,Chen03,Chen03b,Gomez04}. 
The helicity is not a positive definite quantity, and as a result both 
signs of the transfer can indicate a direct cascade depending on the 
sign of the helicity itself. Moreover, as we study smaller scales and the 
relative helicity decreases, both signs have more predominance 
increasing the noise observed in $T_H(K,Q)$.

\section{Conclusions}

We start by summarizing our results. The energy transfer, triadic 
interactions, and statistical flow properties were studied in 
several high resolution DNS with periodic boundaries. The spatial 
resolution ranged from $256^3$ to $1024^3$. The Reynolds number based 
on the integral scale spanned values from $R_e \approx 275 $ to $6200$, 
while the Taylor-based Reynolds number varied between 
$R_\lambda \approx 230$ and $1100$. Three forcing functions were used, 
two coherent functions with zero and maximum net helicity respectively, 
and a random forcing with a short correlation time. The energy injection 
scale was also varied to study its impact on the energy transfer.

Most statistical studies were done for Run III, a high resolution 
simulation with $R_\lambda \approx 800$ and TG forcing. The forcing and 
the resulting flow have spatial symmetries that allow us to identify 
planes with strong and weak large scale shear easily. While in the 
whole domain the standard results were reobtained (e.g. exponential 
and stretched exponential tails in the pdfs of velocity increments, 
and anomalous scaling of the structure functions), when studying 
individual regions a correlation between large scale shear and small 
scale gradients was found. In regions of strong shear, the tails of 
the pdfs of velocity increments are stronger than in regions of weak 
shear, even at scales as small as twice the dissipation scale. Also, 
structure functions show a slightly larger slope in regions of strong 
large scale shear, i.e. a larger departure from a Kolmogorov $p/3$ self-similar scaling.

A correlation between stronger tails in the pdfs of velocity 
increments and the presence of vortex tubes has already been observed 
in different decompositions of the flow \cite{Levi01,Farge01}. Here, 
we observed a correlation between these quantities and the large scale 
shear. The correlation was observed to be persistent even at scales as 
small as the dissipation scale. Small differences were also observed 
in the anomalous scaling of the structure functions for TG and ABC 
forcing. However, it is unclear whether this is related with 
non-universal effects associated with interactions with the large scale 
forcing, or with the use of the ESS hypothesis. For individual vortex 
tubes in several flows, we verified that the flow inside and surrounding 
the vortex tube is helical, as found before 
\cite{Tsinober83,Moffatt85,Moffatt86,Levich87,Farge01}. Note that the 
development of helical structures in a turbulent flow can lead to 
the depletion on nonlinearity and a quenching of local interactions 
\cite{Moffatt92,Tsinober}.

Concerning the energy transfer and triadic interactions, we confirmed 
for several forcing functions that the cascade of energy is local between 
Fourier shells, although it is strongly mediated by individual triadic 
interactions which are nonlocal in nature. As a result, the energy cascades 
from one shell to the next with a fixed step proportional to the forcing 
wavenumber $k_F$. This effect was observed even in simulations using 
random forcing with a correlation time one order of magnitude smaller 
than the large scale turnover time. No qualitative differences have been 
observed as the Reynolds number was changed.

However, the degree of nonlocality observed depends on the quantity 
studied. A hierarchy is found in the relative amplitude of nonlocal 
effects in triadic interactions, the shell-to-shell energy transfer, 
and the energy flux. Triadic interactions are dominated by interactions 
with the large scale flow. As more modes are summed to define the 
shell-to-shell transfer and the flux, the larger population of small 
scale modes starts dominating, and in the $1024^3$ simulations the 
large scale flow is responsible for only $\sim 20\%$ of the total 
flux.

An increase of the relative contribution of nonlocal interactions 
(with the large scale flow and with modes outside the octave band studied) 
in the total flux is observed as the dissipative range is reached. This 
result is in good agreement with claims that the bottleneck effect is due 
to the quenching of local interactions at the end of the inertial range 
because of the presence of a cut-off in wavenumbers 
\cite{Herring82,Lohse95,Martinez97}.

It is worth noting that the local energy cascade through nonlocal 
interactions, or non negligible nonlocal interactions, has been observed 
in the past in simulations at lower Reynolds numbers (see e.g. Refs. 
\cite{Domaradzki90,Okhitani92,Yeung95,Zhou96}). Our results confirm 
the presence of interactions between disparate scales in a turbulent flow 
at much larger Reynolds numbers, and for a variety of forcing functions 
and forcing scales. The results discussed here also shed some light on 
the controversy in the literature about the relevance of the nonlocal 
interactions. It has been claimed that interactions are local if 
a different measure for the locality is introduced \cite{Zhou93,Zhou93b}, 
or if wavelets or a binning of Fourier space in octaves is used 
\cite{Kishida99} (note that wavelets naturally introduce a binning in 
octaves of the spectral space). The hierarchy found in the different 
transfer functions is the reason for this apparent inconsistency 
between previous results. As more modes are summed (e.g. to define 
octaves in spectral space, or to define our partial fluxes $\Pi_P$), 
the small scales overcome the triadic interactions with the large scale 
flow, and local interactions give the largest contribution. This is also 
in agreement with recent theoretical results about the locality of the 
energy flux \cite{Eyink05}, or the locality of the shell-to-shell energy 
transfer in Fourier octave bands \cite{Eyink94,Verma05}.

The fact that in simulations with $R_\lambda \sim 1000$ most of 
the flux is due to local interactions, does not preclude however the 
existence of strong nonlocal interactions with the large scale flow at 
the triadic interaction level, and it should be kept in mind that even 
at large values of $R_\lambda$ these interactions are responsible for 
a non-negligible fraction of the total flux. The presence of nonlocal 
interactions are a deviation from the standard hypothesis often 
associated with Kolmogorov (K41) theory \cite{Kolmogorov41}, 
and can possibly explain departures from self-similar
models of the higher order structure functions and 
controversial results observed in experiments, such as 
the slower than predicted recovery of isotropy in the small scales 
\cite{Shen00}.

Similar results were obtained for the helicity transfer. The 
injection of net helicity in the large scales breaks down the mirror 
symmetry in the flow. While confirming the direct cascade of helicity 
\cite{Brissaud73,Andre77,Borue97,Chen03,Chen03b,Gomez04}, 
we also found a slower than 
expected recovery of the symmetries in the small scales, with an 
excess of relative helicity at small scales compared with the $k^{-1}$ 
expected drop. As in the energy cascade, the cascade of helicity takes 
place in fixed steps proportional to the forcing scale, indicating 
strong nonlocal triadic interactions.

Nonlocal interactions can also be responsible for observed 
departures from universality. In \cite{Alexakis05b} it was shown with 
a simple dimensional argument how nonlocal interactions can still 
be consistent with a $k^{-5/3}$ energy spectrum. Vortex tube stretching 
by the large scale flow plays a significant role, and the argument 
has points in common with multifractal models of intermittency, such 
as the $\beta$ model (see e.g. \cite{Frisch}). As previously mentioned, 
in this work we showed evidence of a positive correlation of large scale 
shear and small scale gradients, both in pdfs of velocity increments and 
in structure functions. Ref. \cite{Laval01} showed that the anomalous 
scaling of the structure functions is reduced when nonlocal interactions 
with the large scale flow are artificially suppressed in a simulation. 
Both results suggest that departures from K41 theory can be associated with 
the imprint of the large scale forcing in turbulence.

The modern language used in turbulence is related to a great extent 
to the K41 theory developed for the isotropic and homogeneous case, with 
the assumption that such fundamental symmetries of the equations would 
be recovered in the small scales even when broken (e.g. through external 
forcing) in the large scales. To apply this theory to real turbulent 
flows, the idea that small 
scales restore isotropy and homogeneity is based strongly on the assumption 
of local interactions between scales. The persistence of 
anisotropies and other deviations from K41 observed in experiments and 
simulations have been associated in this work with the presence of strong 
non-local triadic interactions. At this point, the reader could ask 
how much of the edifice of homogeneous and isotropic turbulence remains.

One of the assumptions in K41 theory is that the properties of the 
inertial range are universal. This is directly related to the assumption 
of local interactions. Since eddies in the inertial range only interact 
with eddies of similar size, as the energy cascades through the inertial 
range a self-similar solution is obtained. The fraction of the flux due to 
non-local interactions, found to be $\sim 20\%$ here,  can 
be interpreted as a subleading contribution to the local flux. 
Deviations from K41 theory are well documented and several theories have been 
proposed to explain them (see e.g. 
\cite{Kolmogorov62,She94,Frisch,Lesieur,Tsinober}). 
They have also been derived theoretically for the passive scalar 
in the context of the so-called Kraichnan model \cite{Kraichnan68}.
Several of these corrections were associated with viscous effects for 
turbulence at finite Reynolds number. What our results show is that 
these deviations can come as well from interactions with the large 
scale flow. In this context, simulations at higher resolution could help 
to study the scaling of the non-local flux with the Reynolds number.

It is known that in the Kraichnan model only prefactors of the scaling 
laws depend on the large-scale forcing (i.e. the exponents are universal: 
independent of the forcing). However, in the case of the Navier-Stokes 
equations, this has not been proved yet since the hierarchy of equations 
for the $n$-point correlation functions is not closed. Then, inertial 
range solutions with exponents that depend on the forcing 
may exist.

In the case of realistic turbulent flows as encountered in astrophysics 
and geophysics, the presence of strong non-local 
interactions can lead to the persistence of anisotropies in the small 
scales. In this case, the anisotropic effects question the applicability 
of the theory of isotropic and homogeneous turbulence to real flows (at 
least at Reynolds numbers comparable to the ones studied in this work). 
The properties of the large scale flow can be more important than 
expected to shape the small scales. A systematic study of the isotropic 
and anisotropic contributions to the scaling 
\cite{Biferale01a,Biferale01b,Biferale02} can be a first step to recognize 
universal (and non-universal) features in these cases.

Finally, as previously indicated in \cite{Alexakis05b}, the existence of
non-negligible nonlocal interactions in a variety of turbulent flows
gives support to models involving as an essential agent of the nonlinear
energy transfer the distortion of turbulent eddies by a large-scale
flow (e.g. as in rapid distortion theory and its variants
\cite{Laval01,Dubrulle04}, or as in the Lagrangian averaged Navier-Stokes 
equations \cite{Holm98,Chen98} where the turbulent flow interacts with a 
smooth velocity field). We believe that the understanding of scale 
interactions 
in turbulence can lead to the development of a new generation of subgrid 
models, beyond the usual hypothesis of locality done in most Large Eddy 
Simulations.

\begin{acknowledgments}
The authors would like to express their gratitude to J.R. Herring for 
valuable discussions and his careful reading of the manuscript. The 
authors also acknowledge discussions with L. Biferale and S. Kurien.
Computer time was provided by NCAR and by the National Science Foundation 
Terascale Computing System at the Pittsburgh Supercomputing Center. The 
NSF grant CMG-0327888 at NCAR supported this work in part and is gratefully 
acknowledged. Three-dimensional visualizations of the flows were done 
using VAPoR, a software for interactive visualization and analysis of 
terascale datasets \cite{vapor}. The authors are grateful to A. Norton 
and J. Clyne (SCD/CISL) for help with the visualizations.
\end{acknowledgments}

\bibliography{ms}

\begin{thebibliography}{67}
\expandafter\ifx\csname natexlab\endcsname\relax\def\natexlab#1{#1}\fi
\expandafter\ifx\csname bibnamefont\endcsname\relax
  \def\bibnamefont#1{#1}\fi
\expandafter\ifx\csname bibfnamefont\endcsname\relax
  \def\bibfnamefont#1{#1}\fi
\expandafter\ifx\csname citenamefont\endcsname\relax
  \def\citenamefont#1{#1}\fi
\expandafter\ifx\csname url\endcsname\relax
  \def\url#1{\texttt{#1}}\fi
\expandafter\ifx\csname urlprefix\endcsname\relax\def\urlprefix{URL }\fi
\providecommand{\bibinfo}[2]{#2}
\providecommand{\eprint}[2][]{\url{#2}}

\bibitem[{\citenamefont{Domaradzki}(1988)}]{Domaradzki88}
\bibinfo{author}{\bibfnamefont{J.~A.} \bibnamefont{Domaradzki}},
  \bibinfo{journal}{Phys. \ Fluids} \textbf{\bibinfo{volume}{31}},
  \bibinfo{pages}{2747} (\bibinfo{year}{1988}).

\bibitem[{\citenamefont{Domaradzki and Rogallo}(1990)}]{Domaradzki90}
\bibinfo{author}{\bibfnamefont{J.~A.} \bibnamefont{Domaradzki}}
  \bibnamefont{and} \bibinfo{author}{\bibfnamefont{R.~S.}
  \bibnamefont{Rogallo}}, \bibinfo{journal}{Phys.\ Fluids A}
  \textbf{\bibinfo{volume}{2}}, \bibinfo{pages}{413} (\bibinfo{year}{1990}).

\bibitem[{\citenamefont{Kerr}(1990)}]{Kerr90}
\bibinfo{author}{\bibfnamefont{R.~M.} \bibnamefont{Kerr}},
  \bibinfo{journal}{J.\ Fluid Mech.} \textbf{\bibinfo{volume}{211}},
  \bibinfo{pages}{309} (\bibinfo{year}{1990}).

\bibitem[{\citenamefont{Yeung and Brasseur}(1991)}]{Yeung91}
\bibinfo{author}{\bibfnamefont{P.~K.} \bibnamefont{Yeung}} \bibnamefont{and}
  \bibinfo{author}{\bibfnamefont{J.~G.} \bibnamefont{Brasseur}},
  \bibinfo{journal}{Phys.\ Fluids A} \textbf{\bibinfo{volume}{3}},
  \bibinfo{pages}{884} (\bibinfo{year}{1991}).

\bibitem[{\citenamefont{Ohkitani and Kida}(1992)}]{Okhitani92}
\bibinfo{author}{\bibfnamefont{K.}~\bibnamefont{Ohkitani}} \bibnamefont{and}
  \bibinfo{author}{\bibfnamefont{S.}~\bibnamefont{Kida}},
  \bibinfo{journal}{Phys.\ Fluids A} \textbf{\bibinfo{volume}{4}},
  \bibinfo{pages}{794} (\bibinfo{year}{1992}).

\bibitem[{\citenamefont{Zhou}(1993{\natexlab{a}})}]{Zhou93}
\bibinfo{author}{\bibfnamefont{Y.}~\bibnamefont{Zhou}},
  \bibinfo{journal}{Phys.\ Fluids A} \textbf{\bibinfo{volume}{5}},
  \bibinfo{pages}{1092} (\bibinfo{year}{1993}{\natexlab{a}}).

\bibitem[{\citenamefont{Zhou}(1993{\natexlab{b}})}]{Zhou93b}
\bibinfo{author}{\bibfnamefont{Y.}~\bibnamefont{Zhou}},
  \bibinfo{journal}{Phys.\ Fluids A} \textbf{\bibinfo{volume}{5}},
  \bibinfo{pages}{2511} (\bibinfo{year}{1993}{\natexlab{b}}).

\bibitem[{\citenamefont{Brasseur and Wei}(1994)}]{Brasseur94}
\bibinfo{author}{\bibfnamefont{J.~G.} \bibnamefont{Brasseur}} \bibnamefont{and}
  \bibinfo{author}{\bibfnamefont{C.~H.} \bibnamefont{Wei}},
  \bibinfo{journal}{Phys.\ Fluids} \textbf{\bibinfo{volume}{6}},
  \bibinfo{pages}{842} (\bibinfo{year}{1994}).

\bibitem[{\citenamefont{Yeung et~al.}(1995)\citenamefont{Yeung, Brasseur, and
  Wang}}]{Yeung95}
\bibinfo{author}{\bibfnamefont{P.~K.} \bibnamefont{Yeung}},
  \bibinfo{author}{\bibfnamefont{J.~G.} \bibnamefont{Brasseur}},
  \bibnamefont{and} \bibinfo{author}{\bibfnamefont{Q.}~\bibnamefont{Wang}},
  \bibinfo{journal}{J.\ Fluid Mech.} \textbf{\bibinfo{volume}{283}},
  \bibinfo{pages}{43} (\bibinfo{year}{1995}).

\bibitem[{\citenamefont{Zhou et~al.}(1996)\citenamefont{Zhou, Yeung, and
  Brasseur}}]{Zhou96}
\bibinfo{author}{\bibfnamefont{Y.}~\bibnamefont{Zhou}},
  \bibinfo{author}{\bibfnamefont{P.~K.} \bibnamefont{Yeung}}, \bibnamefont{and}
  \bibinfo{author}{\bibfnamefont{J.~G.} \bibnamefont{Brasseur}},
  \bibinfo{journal}{Phys.\ Rev.\ E} \textbf{\bibinfo{volume}{53}},
  \bibinfo{pages}{1261} (\bibinfo{year}{1996}).

\bibitem[{\citenamefont{Kishida et~al.}(1999)\citenamefont{Kishida, Araki,
  Kishiba, and Suzuki}}]{Kishida99}
\bibinfo{author}{\bibfnamefont{K.}~\bibnamefont{Kishida}},
  \bibinfo{author}{\bibfnamefont{K.}~\bibnamefont{Araki}},
  \bibinfo{author}{\bibfnamefont{S.}~\bibnamefont{Kishiba}}, \bibnamefont{and}
  \bibinfo{author}{\bibfnamefont{K.}~\bibnamefont{Suzuki}},
  \bibinfo{journal}{Phys.\ Rev.\ Lett.} \textbf{\bibinfo{volume}{83}},
  \bibinfo{pages}{5487} (\bibinfo{year}{1999}).

\bibitem[{\citenamefont{Wiltse and Glezer}(1993)}]{Wiltse93}
\bibinfo{author}{\bibfnamefont{J.~M.} \bibnamefont{Wiltse}} \bibnamefont{and}
  \bibinfo{author}{\bibfnamefont{A.}~\bibnamefont{Glezer}},
  \bibinfo{journal}{J.\ Fluid Mech.} \textbf{\bibinfo{volume}{249}},
  \bibinfo{pages}{261} (\bibinfo{year}{1993}).

\bibitem[{\citenamefont{Wiltse and Glezer}(1998)}]{Wiltse98}
\bibinfo{author}{\bibfnamefont{J.~M.} \bibnamefont{Wiltse}} \bibnamefont{and}
  \bibinfo{author}{\bibfnamefont{A.}~\bibnamefont{Glezer}},
  \bibinfo{journal}{Phys.\ Fluids} \textbf{\bibinfo{volume}{10}},
  \bibinfo{pages}{2026} (\bibinfo{year}{1998}).

\bibitem[{\citenamefont{Carlier et~al.}(2001)\citenamefont{Carlier, Laval, and
  Stanislas}}]{Carlier01}
\bibinfo{author}{\bibfnamefont{J.}~\bibnamefont{Carlier}},
  \bibinfo{author}{\bibfnamefont{J.~P.} \bibnamefont{Laval}}, \bibnamefont{and}
  \bibinfo{author}{\bibfnamefont{M.}~\bibnamefont{Stanislas}},
  \bibinfo{journal}{Compt.\ Rend.\ de l'Academ.\ des Sci.\ Ser. II}
  \textbf{\bibinfo{volume}{329}}, \bibinfo{pages}{35} (\bibinfo{year}{2001}).

\bibitem[{\citenamefont{Sreenivasan and Antonia}(1997)}]{Sreenivasan97}
\bibinfo{author}{\bibfnamefont{K.~R.} \bibnamefont{Sreenivasan}}
  \bibnamefont{and} \bibinfo{author}{\bibfnamefont{R.~A.}
  \bibnamefont{Antonia}}, \bibinfo{journal}{Annu.\ Rev.\ Fluid Mech.} pp.
  \bibinfo{pages}{437--472} (\bibinfo{year}{1997}).

\bibitem[{\citenamefont{Shen and Warhaft}(2000)}]{Shen00}
\bibinfo{author}{\bibfnamefont{X.}~\bibnamefont{Shen}} \bibnamefont{and}
  \bibinfo{author}{\bibfnamefont{Z.}~\bibnamefont{Warhaft}},
  \bibinfo{journal}{Phys.\ Fluids} \textbf{\bibinfo{volume}{12}},
  \bibinfo{pages}{2976} (\bibinfo{year}{2000}).

\bibitem[{\citenamefont{Stewart}(1969)}]{Stewart69}
\bibinfo{author}{\bibfnamefont{R.~W.} \bibnamefont{Stewart}},
  \bibinfo{journal}{Radio Sci.} \textbf{\bibinfo{volume}{4}},
  \bibinfo{pages}{1269} (\bibinfo{year}{1969}).

\bibitem[{\citenamefont{Biferale and Toschi}(2001)}]{Biferale01a}
\bibinfo{author}{\bibfnamefont{L.}~\bibnamefont{Biferale}} \bibnamefont{and}
  \bibinfo{author}{\bibfnamefont{F.}~\bibnamefont{Toschi}},
  \bibinfo{journal}{Phys.\ Rev.\ Lett.} \textbf{\bibinfo{volume}{86}},
  \bibinfo{pages}{4831} (\bibinfo{year}{2001}).

\bibitem[{\citenamefont{Laval et~al.}(2001)\citenamefont{Laval, Dubrulle, and
  Nazarenko}}]{Laval01}
\bibinfo{author}{\bibfnamefont{J.-P.} \bibnamefont{Laval}},
  \bibinfo{author}{\bibfnamefont{B.}~\bibnamefont{Dubrulle}}, \bibnamefont{and}
  \bibinfo{author}{\bibfnamefont{S.}~\bibnamefont{Nazarenko}},
  \bibinfo{journal}{Phys.\ Fluids} \textbf{\bibinfo{volume}{13}},
  \bibinfo{pages}{1995} (\bibinfo{year}{2001}).

\bibitem[{\citenamefont{Alexakis
  et~al.}(2005{\natexlab{a}})\citenamefont{Alexakis, Mininni, and
  Pouquet}}]{Alexakis05b}
\bibinfo{author}{\bibfnamefont{A.}~\bibnamefont{Alexakis}},
  \bibinfo{author}{\bibfnamefont{P.~D.} \bibnamefont{Mininni}},
  \bibnamefont{and} \bibinfo{author}{\bibfnamefont{A.}~\bibnamefont{Pouquet}},
  \bibinfo{journal}{Phys.\ Rev.\ Lett.} \textbf{\bibinfo{volume}{95}},
  \bibinfo{pages}{264503} (\bibinfo{year}{2005}{\natexlab{a}}).

\bibitem[{\citenamefont{G\'omez
  et~al.}(2005{\natexlab{a}})\citenamefont{G\'omez, Mininni, and
  Dmitruk}}]{Gomez05a}
\bibinfo{author}{\bibfnamefont{D.~O.} \bibnamefont{G\'omez}},
  \bibinfo{author}{\bibfnamefont{P.~D.} \bibnamefont{Mininni}},
  \bibnamefont{and} \bibinfo{author}{\bibfnamefont{P.}~\bibnamefont{Dmitruk}},
  \bibinfo{journal}{Adv.\ Sp.\ Res.} \textbf{\bibinfo{volume}{35}},
  \bibinfo{pages}{899} (\bibinfo{year}{2005}{\natexlab{a}}).

\bibitem[{\citenamefont{G\'omez
  et~al.}(2005{\natexlab{b}})\citenamefont{G\'omez, Mininni, and
  Dmitruk}}]{Gomez05b}
\bibinfo{author}{\bibfnamefont{D.~O.} \bibnamefont{G\'omez}},
  \bibinfo{author}{\bibfnamefont{P.~D.} \bibnamefont{Mininni}},
  \bibnamefont{and} \bibinfo{author}{\bibfnamefont{P.}~\bibnamefont{Dmitruk}},
  \bibinfo{journal}{Phys.\ Scripta} \textbf{\bibinfo{volume}{T116}},
  \bibinfo{pages}{123} (\bibinfo{year}{2005}{\natexlab{b}}).

\bibitem[{\citenamefont{Taylor and Green}(1937)}]{Taylor37}
\bibinfo{author}{\bibfnamefont{G.~I.} \bibnamefont{Taylor}} \bibnamefont{and}
  \bibinfo{author}{\bibfnamefont{A.~E.} \bibnamefont{Green}},
  \bibinfo{journal}{Proc.\ Roy.\ Soc.\ Lond.\ Ser.\ A}
  \textbf{\bibinfo{volume}{158}}, \bibinfo{pages}{499} (\bibinfo{year}{1937}).

\bibitem[{\citenamefont{Podvigina and Pouquet}(1994)}]{Podvigina94}
\bibinfo{author}{\bibfnamefont{O.}~\bibnamefont{Podvigina}} \bibnamefont{and}
  \bibinfo{author}{\bibfnamefont{A.}~\bibnamefont{Pouquet}},
  \bibinfo{journal}{Physica D} \textbf{\bibinfo{volume}{75}},
  \bibinfo{pages}{471} (\bibinfo{year}{1994}).

\bibitem[{\citenamefont{Kraichnan}(1971)}]{Kraichnan71}
\bibinfo{author}{\bibfnamefont{R.~H.} \bibnamefont{Kraichnan}},
  \bibinfo{journal}{J.\ Fluid Mech.} \textbf{\bibinfo{volume}{47}},
  \bibinfo{pages}{525} (\bibinfo{year}{1971}).

\bibitem[{\citenamefont{Lesieur}(1997)}]{Lesieur}
\bibinfo{author}{\bibfnamefont{M.}~\bibnamefont{Lesieur}},
  \emph{\bibinfo{title}{Turbulence in fluids}} (\bibinfo{publisher}{Kluwer
  Acad.\ Press}, \bibinfo{address}{Dordrecht}, \bibinfo{year}{1997}).

\bibitem[{\citenamefont{Alexakis
  et~al.}(2005{\natexlab{b}})\citenamefont{Alexakis, Mininni, and
  Pouquet}}]{Alexakis05}
\bibinfo{author}{\bibfnamefont{A.}~\bibnamefont{Alexakis}},
  \bibinfo{author}{\bibfnamefont{P.~D.} \bibnamefont{Mininni}},
  \bibnamefont{and} \bibinfo{author}{\bibfnamefont{A.}~\bibnamefont{Pouquet}},
  \bibinfo{journal}{Phys.\ Rev.\ E} \textbf{\bibinfo{volume}{72}},
  \bibinfo{pages}{046301} (\bibinfo{year}{2005}{\natexlab{b}}).

\bibitem[{\citenamefont{Frisch}(1995)}]{Frisch}
\bibinfo{author}{\bibfnamefont{U.}~\bibnamefont{Frisch}},
  \emph{\bibinfo{title}{Turbulence: the legacy of A.N. Kolmogorov}}
  (\bibinfo{publisher}{Cambridge Univ.\ Press}, \bibinfo{address}{Cambridge},
  \bibinfo{year}{1995}).

\bibitem[{\citenamefont{Alexakis et~al.}(2006)\citenamefont{Alexakis, Mininni,
  and Pouquet}}]{Alexakis06}
\bibinfo{author}{\bibfnamefont{A.}~\bibnamefont{Alexakis}},
  \bibinfo{author}{\bibfnamefont{P.~D.} \bibnamefont{Mininni}},
  \bibnamefont{and} \bibinfo{author}{\bibfnamefont{A.}~\bibnamefont{Pouquet}},
  \bibinfo{journal}{Astrophys.\ J.}  (\bibinfo{year}{2006}), \bibinfo{note}{in
  press}, \eprint{physics/0509069}.

\bibitem[{\citenamefont{Tsinober and Levich}(1983)}]{Tsinober83}
\bibinfo{author}{\bibfnamefont{A.}~\bibnamefont{Tsinober}} \bibnamefont{and}
  \bibinfo{author}{\bibfnamefont{E.}~\bibnamefont{Levich}},
  \bibinfo{journal}{Phys.\ Lett.\ A} \textbf{\bibinfo{volume}{99}},
  \bibinfo{pages}{321} (\bibinfo{year}{1983}).

\bibitem[{\citenamefont{Moffatt}(1985)}]{Moffatt85}
\bibinfo{author}{\bibfnamefont{H.~K.} \bibnamefont{Moffatt}},
  \bibinfo{journal}{J.\ Fluid Mech.} \textbf{\bibinfo{volume}{159}},
  \bibinfo{pages}{359} (\bibinfo{year}{1985}).

\bibitem[{\citenamefont{Moffatt}(1986)}]{Moffatt86}
\bibinfo{author}{\bibfnamefont{H.~K.} \bibnamefont{Moffatt}},
  \bibinfo{journal}{J.\ Fluid Mech.} \textbf{\bibinfo{volume}{166}},
  \bibinfo{pages}{359} (\bibinfo{year}{1986}).

\bibitem[{\citenamefont{Levich}(1987)}]{Levich87}
\bibinfo{author}{\bibfnamefont{E.}~\bibnamefont{Levich}},
  \bibinfo{journal}{Phys.\ Rep.} \textbf{\bibinfo{volume}{151}},
  \bibinfo{pages}{129} (\bibinfo{year}{1987}).

\bibitem[{\citenamefont{Farge et~al.}(2001)\citenamefont{Farge, Pellegrino, and
  Schneider}}]{Farge01}
\bibinfo{author}{\bibfnamefont{M.}~\bibnamefont{Farge}},
  \bibinfo{author}{\bibfnamefont{G.}~\bibnamefont{Pellegrino}},
  \bibnamefont{and}
  \bibinfo{author}{\bibfnamefont{K.}~\bibnamefont{Schneider}},
  \bibinfo{journal}{Phys.\ Rev.\ Lett.} \textbf{\bibinfo{volume}{87}},
  \bibinfo{pages}{054501} (\bibinfo{year}{2001}).

\bibitem[{\citenamefont{Moffatt and Tsinober}(1992)}]{Moffatt92}
\bibinfo{author}{\bibfnamefont{H.~K.} \bibnamefont{Moffatt}} \bibnamefont{and}
  \bibinfo{author}{\bibfnamefont{A.}~\bibnamefont{Tsinober}},
  \bibinfo{journal}{Annu.\ Rev.\ Fluid Mech.} \textbf{\bibinfo{volume}{24}},
  \bibinfo{pages}{281} (\bibinfo{year}{1992}).

\bibitem[{\citenamefont{Tsinober}(2001)}]{Tsinober}
\bibinfo{author}{\bibfnamefont{A.}~\bibnamefont{Tsinober}},
  \emph{\bibinfo{title}{An informal introduction to turbulence}}
  (\bibinfo{publisher}{Kluwer Acad.\ Press}, \bibinfo{address}{Dordrecht},
  \bibinfo{year}{2001}).

\bibitem[{\citenamefont{Levi and Montgomery}(2001)}]{Levi01}
\bibinfo{author}{\bibfnamefont{T.~S.} \bibnamefont{Levi}} \bibnamefont{and}
  \bibinfo{author}{\bibfnamefont{D.~C.} \bibnamefont{Montgomery}},
  \bibinfo{journal}{Phys.\ Rev.\ E} \textbf{\bibinfo{volume}{63}},
  \bibinfo{pages}{056311} (\bibinfo{year}{2001}).

\bibitem[{\citenamefont{Herring et~al.}(1982)\citenamefont{Herring, Schertzer,
  Lesieur, Newman, Chollet, and Larcheveque}}]{Herring82}
\bibinfo{author}{\bibfnamefont{J.~R.} \bibnamefont{Herring}},
  \bibinfo{author}{\bibfnamefont{D.}~\bibnamefont{Schertzer}},
  \bibinfo{author}{\bibfnamefont{M.}~\bibnamefont{Lesieur}},
  \bibinfo{author}{\bibfnamefont{G.~R.} \bibnamefont{Newman}},
  \bibinfo{author}{\bibfnamefont{J.~P.} \bibnamefont{Chollet}},
  \bibnamefont{and}
  \bibinfo{author}{\bibfnamefont{M.}~\bibnamefont{Larcheveque}},
  \bibinfo{journal}{J.\ Fluid Mech} \textbf{\bibinfo{volume}{124}},
  \bibinfo{pages}{411} (\bibinfo{year}{1982}).

\bibitem[{\citenamefont{Falkovich}(1994)}]{Falkovich94}
\bibinfo{author}{\bibfnamefont{G.}~\bibnamefont{Falkovich}},
  \bibinfo{journal}{Phys.\ Fluids} \textbf{\bibinfo{volume}{6}},
  \bibinfo{pages}{1411} (\bibinfo{year}{1994}).

\bibitem[{\citenamefont{Lohse and M\"uller-Groeling}(1995)}]{Lohse95}
\bibinfo{author}{\bibfnamefont{D.}~\bibnamefont{Lohse}} \bibnamefont{and}
  \bibinfo{author}{\bibfnamefont{A.}~\bibnamefont{M\"uller-Groeling}},
  \bibinfo{journal}{Phys.\ Rev.\ Lett.} \textbf{\bibinfo{volume}{74}},
  \bibinfo{pages}{1747} (\bibinfo{year}{1995}).

\bibitem[{\citenamefont{Mart\'{\i}nez et~al.}(1997)\citenamefont{Mart\'{\i}nez,
  Chen, Doolen, Kraichnan, Wang, and Zhou}}]{Martinez97}
\bibinfo{author}{\bibfnamefont{D.~O.} \bibnamefont{Mart\'{\i}nez}},
  \bibinfo{author}{\bibfnamefont{S.}~\bibnamefont{Chen}},
  \bibinfo{author}{\bibfnamefont{G.~D.} \bibnamefont{Doolen}},
  \bibinfo{author}{\bibfnamefont{R.~H.} \bibnamefont{Kraichnan}},
  \bibinfo{author}{\bibfnamefont{L.-P.} \bibnamefont{Wang}}, \bibnamefont{and}
  \bibinfo{author}{\bibfnamefont{Y.}~\bibnamefont{Zhou}}, \bibinfo{journal}{J.\
  Plasma Phys.} \textbf{\bibinfo{volume}{57}}, \bibinfo{pages}{195}
  (\bibinfo{year}{1997}).

\bibitem[{\citenamefont{Childress and Gilbert}(1995)}]{Childress}
\bibinfo{author}{\bibfnamefont{S.}~\bibnamefont{Childress}} \bibnamefont{and}
  \bibinfo{author}{\bibfnamefont{A.~D.} \bibnamefont{Gilbert}},
  \emph{\bibinfo{title}{Stretch, twist, fold: the fast dynamo}}
  (\bibinfo{publisher}{Springer-Verlag}, \bibinfo{address}{Berlin},
  \bibinfo{year}{1995}).

\bibitem[{\citenamefont{Brissaud et~al.}(1973)\citenamefont{Brissaud, Frisch,
  Leorat, Lesieur, and Mazure}}]{Brissaud73}
\bibinfo{author}{\bibfnamefont{A.}~\bibnamefont{Brissaud}},
  \bibinfo{author}{\bibfnamefont{U.}~\bibnamefont{Frisch}},
  \bibinfo{author}{\bibfnamefont{J.}~\bibnamefont{Leorat}},
  \bibinfo{author}{\bibfnamefont{M.}~\bibnamefont{Lesieur}}, \bibnamefont{and}
  \bibinfo{author}{\bibfnamefont{A.}~\bibnamefont{Mazure}},
  \bibinfo{journal}{Phys.\ Fluids} \textbf{\bibinfo{volume}{16}},
  \bibinfo{pages}{1366} (\bibinfo{year}{1973}).

\bibitem[{\citenamefont{Borue and Orszag}(1997)}]{Borue97}
\bibinfo{author}{\bibfnamefont{V.}~\bibnamefont{Borue}} \bibnamefont{and}
  \bibinfo{author}{\bibfnamefont{S.~A.} \bibnamefont{Orszag}},
  \bibinfo{journal}{Phys.\ Rev.\ E} \textbf{\bibinfo{volume}{55}},
  \bibinfo{pages}{7005} (\bibinfo{year}{1997}).

\bibitem[{\citenamefont{Chen et~al.}(2003{\natexlab{a}})\citenamefont{Chen,
  Chen, and Eyink}}]{Chen03}
\bibinfo{author}{\bibfnamefont{Q.}~\bibnamefont{Chen}},
  \bibinfo{author}{\bibfnamefont{S.}~\bibnamefont{Chen}}, \bibnamefont{and}
  \bibinfo{author}{\bibfnamefont{G.~L.} \bibnamefont{Eyink}},
  \bibinfo{journal}{Phys.\ Fluids} \textbf{\bibinfo{volume}{15}},
  \bibinfo{pages}{361} (\bibinfo{year}{2003}{\natexlab{a}}).

\bibitem[{\citenamefont{Chen et~al.}(2003{\natexlab{b}})\citenamefont{Chen,
  Chen, Eyink, and Holm}}]{Chen03b}
\bibinfo{author}{\bibfnamefont{Q.}~\bibnamefont{Chen}},
  \bibinfo{author}{\bibfnamefont{S.}~\bibnamefont{Chen}},
  \bibinfo{author}{\bibfnamefont{G.~L.} \bibnamefont{Eyink}}, \bibnamefont{and}
  \bibinfo{author}{\bibfnamefont{D.~D.} \bibnamefont{Holm}},
  \bibinfo{journal}{Phys.\ Rev.\ Lett.} \textbf{\bibinfo{volume}{90}},
  \bibinfo{pages}{214503} (\bibinfo{year}{2003}{\natexlab{b}}).

\bibitem[{\citenamefont{G\'omez and Mininni}(2004)}]{Gomez04}
\bibinfo{author}{\bibfnamefont{D.~O.} \bibnamefont{G\'omez}} \bibnamefont{and}
  \bibinfo{author}{\bibfnamefont{P.~D.} \bibnamefont{Mininni}},
  \bibinfo{journal}{Physica A} \textbf{\bibinfo{volume}{342}},
  \bibinfo{pages}{69} (\bibinfo{year}{2004}).

\bibitem[{\citenamefont{Andr\'e and Lesieur}(1997)}]{Andre77}
\bibinfo{author}{\bibfnamefont{J.~C.} \bibnamefont{Andr\'e}} \bibnamefont{and}
  \bibinfo{author}{\bibfnamefont{M.}~\bibnamefont{Lesieur}},
  \bibinfo{journal}{J.\ Fluid Mech.} \textbf{\bibinfo{volume}{81}},
  \bibinfo{pages}{187} (\bibinfo{year}{1997}).

\bibitem[{\citenamefont{Kolmogorov}(1941)}]{Kolmogorov41}
\bibinfo{author}{\bibfnamefont{A.~N.} \bibnamefont{Kolmogorov}},
  \bibinfo{journal}{Dokl.\ Akad.\ Nauk SSSR} \textbf{\bibinfo{volume}{30}},
  \bibinfo{pages}{9} (\bibinfo{year}{1941}).

\bibitem[{\citenamefont{She and L\'ev\^eque}(1994)}]{She94}
\bibinfo{author}{\bibfnamefont{Z.~S.} \bibnamefont{She}} \bibnamefont{and}
  \bibinfo{author}{\bibfnamefont{E.}~\bibnamefont{L\'ev\^eque}},
  \bibinfo{journal}{Phys.\ Rev.\ Lett.} \textbf{\bibinfo{volume}{72}},
  \bibinfo{pages}{336} (\bibinfo{year}{1994}).

\bibitem[{\citenamefont{Benzi et~al.}(1993{\natexlab{a}})\citenamefont{Benzi,
  Ciliberto, Baudet, Chavarria, and Tripiccione}}]{Benzi93}
\bibinfo{author}{\bibfnamefont{R.}~\bibnamefont{Benzi}},
  \bibinfo{author}{\bibfnamefont{S.}~\bibnamefont{Ciliberto}},
  \bibinfo{author}{\bibfnamefont{C.}~\bibnamefont{Baudet}},
  \bibinfo{author}{\bibfnamefont{G.~R.} \bibnamefont{Chavarria}},
  \bibnamefont{and}
  \bibinfo{author}{\bibfnamefont{R.}~\bibnamefont{Tripiccione}},
  \bibinfo{journal}{Europhys.\ Lett.} \textbf{\bibinfo{volume}{24}},
  \bibinfo{pages}{275} (\bibinfo{year}{1993}{\natexlab{a}}).

\bibitem[{\citenamefont{Benzi et~al.}(1993{\natexlab{b}})\citenamefont{Benzi,
  Ciliberto, Tripiccione, Baudet, Massaioli, and Succi}}]{Benzi93b}
\bibinfo{author}{\bibfnamefont{R.}~\bibnamefont{Benzi}},
  \bibinfo{author}{\bibfnamefont{S.}~\bibnamefont{Ciliberto}},
  \bibinfo{author}{\bibfnamefont{R.}~\bibnamefont{Tripiccione}},
  \bibinfo{author}{\bibfnamefont{C.}~\bibnamefont{Baudet}},
  \bibinfo{author}{\bibfnamefont{F.}~\bibnamefont{Massaioli}},
  \bibnamefont{and} \bibinfo{author}{\bibfnamefont{S.}~\bibnamefont{Succi}},
  \bibinfo{journal}{Phys.\ Rev.\ E} \textbf{\bibinfo{volume}{48}},
  \bibinfo{pages}{R29} (\bibinfo{year}{1993}{\natexlab{b}}).

\bibitem[{\citenamefont{Biferale and Vergassola}(2001)}]{Biferale01b}
\bibinfo{author}{\bibfnamefont{L.}~\bibnamefont{Biferale}} \bibnamefont{and}
  \bibinfo{author}{\bibfnamefont{M.}~\bibnamefont{Vergassola}},
  \bibinfo{journal}{Phys.\ Fluids} \textbf{\bibinfo{volume}{13}},
  \bibinfo{pages}{2139} (\bibinfo{year}{2001}).

\bibitem[{\citenamefont{Biferale et~al.}(2002)\citenamefont{Biferale, Daumont,
  Lanotte, and Toschi}}]{Biferale02}
\bibinfo{author}{\bibfnamefont{L.}~\bibnamefont{Biferale}},
  \bibinfo{author}{\bibfnamefont{I.}~\bibnamefont{Daumont}},
  \bibinfo{author}{\bibfnamefont{A.}~\bibnamefont{Lanotte}}, \bibnamefont{and}
  \bibinfo{author}{\bibfnamefont{F.}~\bibnamefont{Toschi}},
  \bibinfo{journal}{Phys.\ Rev.\ E} \textbf{\bibinfo{volume}{66}},
  \bibinfo{pages}{056306} (\bibinfo{year}{2002}).

\bibitem[{\citenamefont{Waleffe}(1991)}]{Waleffe91}
\bibinfo{author}{\bibfnamefont{F.}~\bibnamefont{Waleffe}},
  \bibinfo{journal}{Phys.\ Fluids A} \textbf{\bibinfo{volume}{4}},
  \bibinfo{pages}{350} (\bibinfo{year}{1991}).

\bibitem[{\citenamefont{Ditlevsen and
  Giuliani}(2001{\natexlab{a}})}]{Ditlevsen01}
\bibinfo{author}{\bibfnamefont{P.~D.} \bibnamefont{Ditlevsen}}
  \bibnamefont{and} \bibinfo{author}{\bibfnamefont{P.}~\bibnamefont{Giuliani}},
  \bibinfo{journal}{Phys.\ Fluids} \textbf{\bibinfo{volume}{13}},
  \bibinfo{pages}{3508} (\bibinfo{year}{2001}{\natexlab{a}}).

\bibitem[{\citenamefont{Ditlevsen and
  Giuliani}(2001{\natexlab{b}})}]{Ditlevsen01b}
\bibinfo{author}{\bibfnamefont{P.~D.} \bibnamefont{Ditlevsen}}
  \bibnamefont{and} \bibinfo{author}{\bibfnamefont{P.}~\bibnamefont{Giuliani}},
  \bibinfo{journal}{Phys.\ Rev.\ E} \textbf{\bibinfo{volume}{63}},
  \bibinfo{pages}{036304} (\bibinfo{year}{2001}{\natexlab{b}}).

\bibitem[{\citenamefont{Kurien et~al.}(2004)\citenamefont{Kurien, Taylor, and
  Matsumoto}}]{Kurien04}
\bibinfo{author}{\bibfnamefont{S.}~\bibnamefont{Kurien}},
  \bibinfo{author}{\bibfnamefont{M.~A.} \bibnamefont{Taylor}},
  \bibnamefont{and}
  \bibinfo{author}{\bibfnamefont{T.}~\bibnamefont{Matsumoto}},
  \bibinfo{journal}{Phys.\ Rev.\ E} \textbf{\bibinfo{volume}{69}},
  \bibinfo{pages}{066313} (\bibinfo{year}{2004}).

\bibitem[{\citenamefont{Eyink}(2005)}]{Eyink05}
\bibinfo{author}{\bibfnamefont{G.~L.} \bibnamefont{Eyink}},
  \bibinfo{journal}{Physica D} \textbf{\bibinfo{volume}{207}},
  \bibinfo{pages}{91} (\bibinfo{year}{2005}).

\bibitem[{\citenamefont{Eyink}(1994)}]{Eyink94}
\bibinfo{author}{\bibfnamefont{G.~L.} \bibnamefont{Eyink}},
  \bibinfo{journal}{Physica D} \textbf{\bibinfo{volume}{78}},
  \bibinfo{pages}{222} (\bibinfo{year}{1994}).

\bibitem[{\citenamefont{Verma et~al.}(2005)\citenamefont{Verma, Ayyer,
  Debliquy, Kumar, and Chandra}}]{Verma05}
\bibinfo{author}{\bibfnamefont{M.~K.} \bibnamefont{Verma}},
  \bibinfo{author}{\bibfnamefont{A.}~\bibnamefont{Ayyer}},
  \bibinfo{author}{\bibfnamefont{O.}~\bibnamefont{Debliquy}},
  \bibinfo{author}{\bibfnamefont{S.}~\bibnamefont{Kumar}}, \bibnamefont{and}
  \bibinfo{author}{\bibfnamefont{A.~V.} \bibnamefont{Chandra}},
  \bibinfo{journal}{Pramana J.\ Phys.} \textbf{\bibinfo{volume}{65}},
  \bibinfo{pages}{297} (\bibinfo{year}{2005}).

\bibitem[{\citenamefont{Kolmogorov}(1962)}]{Kolmogorov62}
\bibinfo{author}{\bibfnamefont{A.~N.} \bibnamefont{Kolmogorov}},
  \bibinfo{journal}{J.\ Fluid Mech.} \textbf{\bibinfo{volume}{13}},
  \bibinfo{pages}{82} (\bibinfo{year}{1962}).

\bibitem[{\citenamefont{Kraichnan}(1968)}]{Kraichnan68}
\bibinfo{author}{\bibfnamefont{R.~H.} \bibnamefont{Kraichnan}},
  \bibinfo{journal}{Phys.\ Fluids} \textbf{\bibinfo{volume}{11}},
  \bibinfo{pages}{945} (\bibinfo{year}{1968}).

\bibitem[{\citenamefont{Dubrulle et~al.}(2004)\citenamefont{Dubrulle, Laval,
  Nazarenko, and Zaboronski}}]{Dubrulle04}
\bibinfo{author}{\bibfnamefont{B.}~\bibnamefont{Dubrulle}},
  \bibinfo{author}{\bibfnamefont{J.-P.} \bibnamefont{Laval}},
  \bibinfo{author}{\bibfnamefont{S.}~\bibnamefont{Nazarenko}},
  \bibnamefont{and}
  \bibinfo{author}{\bibfnamefont{O.}~\bibnamefont{Zaboronski}},
  \bibinfo{journal}{J.\ Fluid Mech.} \textbf{\bibinfo{volume}{520}},
  \bibinfo{pages}{1} (\bibinfo{year}{2004}).

\bibitem[{\citenamefont{Holm et~al.}(1998)\citenamefont{Holm, Marsden, and
  Ratiu}}]{Holm98}
\bibinfo{author}{\bibfnamefont{D.~D.} \bibnamefont{Holm}},
  \bibinfo{author}{\bibfnamefont{J.~E.} \bibnamefont{Marsden}},
  \bibnamefont{and} \bibinfo{author}{\bibfnamefont{T.~S.} \bibnamefont{Ratiu}},
  \bibinfo{journal}{Phys.\ Rev.\ Lett.} \textbf{\bibinfo{volume}{80}},
  \bibinfo{pages}{4173} (\bibinfo{year}{1998}).

\bibitem[{\citenamefont{Chen et~al.}(1998)\citenamefont{Chen, Holm, Foias,
  Olson, Titi, and Wynne}}]{Chen98}
\bibinfo{author}{\bibfnamefont{S.~Y.} \bibnamefont{Chen}},
  \bibinfo{author}{\bibfnamefont{D.~D.} \bibnamefont{Holm}},
  \bibinfo{author}{\bibfnamefont{C.}~\bibnamefont{Foias}},
  \bibinfo{author}{\bibfnamefont{E.~J.} \bibnamefont{Olson}},
  \bibinfo{author}{\bibfnamefont{E.~S.} \bibnamefont{Titi}}, \bibnamefont{and}
  \bibinfo{author}{\bibfnamefont{S.}~\bibnamefont{Wynne}},
  \bibinfo{journal}{Phys.\ Rev.\ Lett.} pp. \bibinfo{pages}{5338--5341}
  (\bibinfo{year}{1998}).

\bibitem[{\citenamefont{Clyne and Rast}(2005)}]{vapor}
\bibinfo{author}{\bibfnamefont{J.}~\bibnamefont{Clyne}} \bibnamefont{and}
  \bibinfo{author}{\bibfnamefont{M.}~\bibnamefont{Rast}}, in
  \emph{\bibinfo{booktitle}{Visualization and data analysis 2005}}, edited by
  \bibinfo{editor}{\bibfnamefont{R.~F.} \bibnamefont{Erbacher}},
  \bibinfo{editor}{\bibfnamefont{J.~C.} \bibnamefont{Roberts}},
  \bibinfo{editor}{\bibfnamefont{M.~T.} \bibnamefont{Grohn}}, \bibnamefont{and}
  \bibinfo{editor}{\bibfnamefont{K.}~\bibnamefont{Borner}}
  (\bibinfo{publisher}{SPIE}, \bibinfo{address}{Bellingham, Wash.},
  \bibinfo{year}{2005}), pp. \bibinfo{pages}{284--294},
  \bibinfo{note}{http://www.vapor.ucar.edu}.

\end{thebibliography}

\end{document}